\documentclass[journal=jacsat,manuscript=article]{achemso}

\usepackage[version=3]{mhchem} 
\usepackage{xcolor}

\DeclareUnicodeCharacter{2009}{\,}

\author{T. de Ara}
\affiliation{Departamento de F\'\i sica Aplicada and Instituto
Universitario de Materiales de Alicante (IUMA), Universidad de Alicante, Campus de San Vicente del Raspeig, E-03690 Alicante, Spain.}

\author{C. Hsu}
\affiliation{Department of Quantum Nanoscience, Delft University of Technology, Delft 2628CJ, The Netherlands.}

\author{A. Martinez-Garcia}
\affiliation{Departamento de F\'\i sica Aplicada and Instituto
Universitario de Materiales de Alicante (IUMA), Universidad de Alicante, Campus de San Vicente del Raspeig, E-03690 Alicante, Spain.}

\author{B. C. Baciu}
\affiliation{Departamento de Qu\'\i mica Org\'anica and Instituto Universitario de S\'\i ntesis Org\'anica, Universidad de Alicante, Campus de San Vicente del Raspeig, E-03690, Alicante, Spain}

\author{P. J. Bronk}
\affiliation{Departamento de Qu\'\i mica Org\'anica and Instituto Universitario de S\'\i ntesis Org\'anica, Universidad de Alicante, Campus de San Vicente del Raspeig, E-03690, Alicante, Spain}

\author{L. Ornago}
\affiliation{Department of Quantum Nanoscience, Delft University of Technology, Delft 2628CJ, The Netherlands.}

\author{S. van der Poel}
\affiliation{Department of Quantum Nanoscience, Delft University of Technology, Delft 2628CJ, The Netherlands.}

\author{E. B. Lombardi}
\affiliation{Department of Physics, Florida Science Campus,
University of South Africa, Florida Park, Johannesburg 1710, South Africa}

\author{A. Guijarro}
\affiliation{Departamento de Qu\'\i mica Org\'anica and Instituto Universitario de S\'\i ntesis Org\'anica, Universidad de Alicante, Campus de San Vicente del Raspeig, E-03690, Alicante, Spain}

\author{C. Sabater}
\affiliation{Departamento de F\'\i sica Aplicada and Instituto
Universitario de Materiales de Alicante (IUMA), Universidad de Alicante, Campus de San Vicente del Raspeig, E-03690 Alicante, Spain.}

\author{C. Untiedt}
\affiliation{Departamento de F\'\i sica Aplicada and Instituto
Universitario de Materiales de Alicante (IUMA), Universidad de Alicante, Campus de San Vicente del Raspeig, E-03690 Alicante, Spain.}

\author{H. S. J. van der Zant}
\affiliation{Department of Quantum Nanoscience, Delft University of Technology, Delft 2628CJ, The Netherlands.}

\email{untiedt@ua.es}

\title{Evidence of an Off-resonant Electronic Transport Mechanism in Helicenes}


\begin{document}

\begin{tocentry}
\includegraphics[width=1\textwidth]{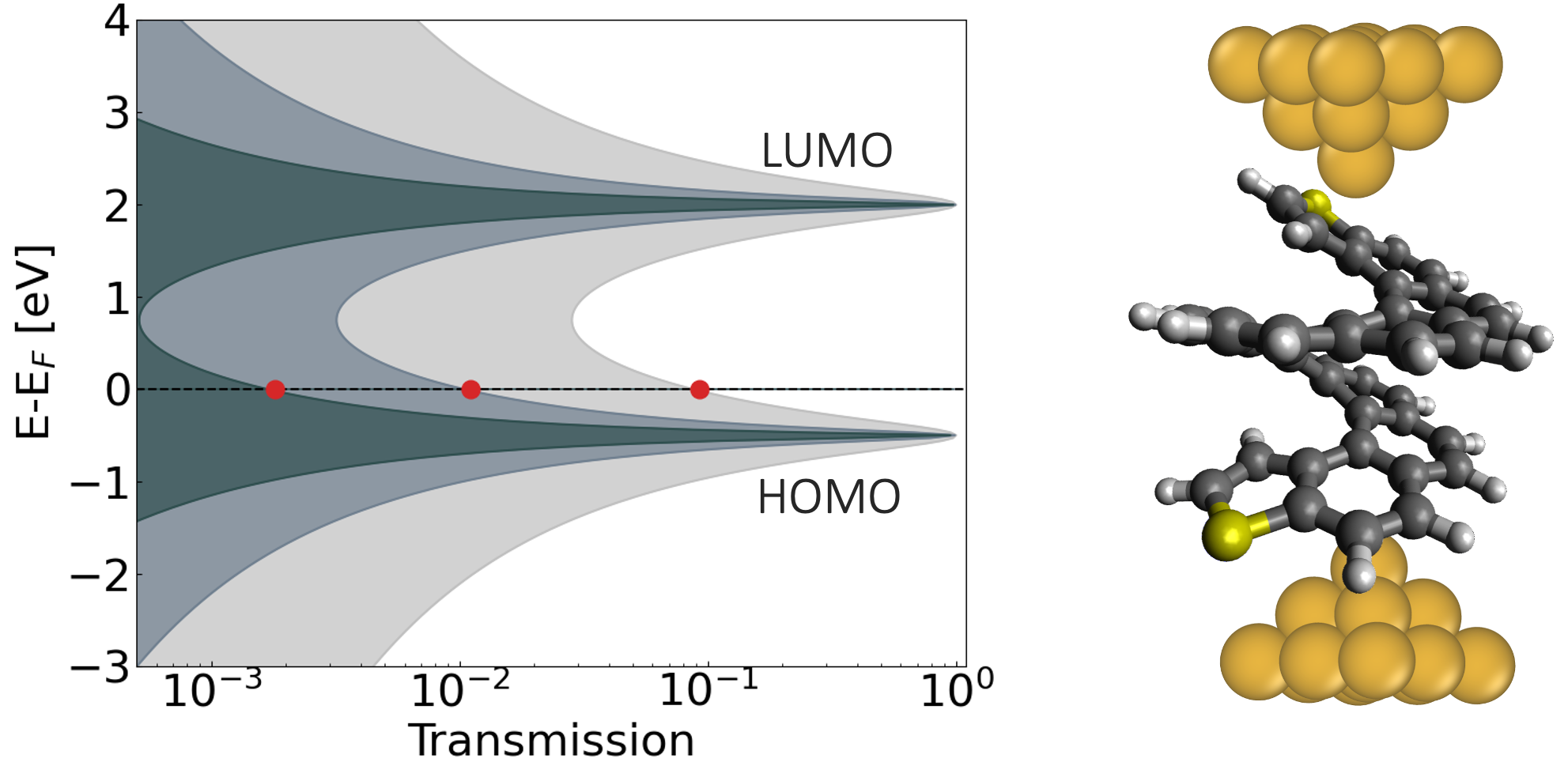}
\end{tocentry}

\begin{abstract}
Helical molecules have been identified as potential candidates for investigating electronic transport, spin filtering, or even piezoelectricity. However, the description of the transport mechanism is not straightforward in single molecular junctions. In this work, we study the electronic transport in break junctions of a series of three helical molecules: dithia[$n$]helicenes, with $n=7, 9, 11$ molecular units, and detail the synthesis of two kinds of dithia[11]helicenes, varying the location of the sulfur atoms. Our experimental study demonstrates low conductance values that remain similar across different biases and molecules. Additionally, we assess the length dependence of the conductance for each helicene, revealing an exponential decay characteristic of off-resonant transport. This behaviour is primarily attributed to the misalignment between the energy levels of the molecule-electrodes system. The length dependence trend described above is supported by \textit{ab initio} calculations, further confirming the off-resonant transport mechanism.
\end{abstract}


The study of electronic transport properties in single molecules has garnered significant attention due to their versatile and programmable structural features. This interest has driven advancements in molecular-scale devices, with the goal of harnessing the unique chemical and physical properties of individual molecules~\cite{singlemoleculeswitch}. A topic that has gained attention in recent years is the research on spin valve-type molecules that do not rely on ferromagnetic electrodes or applied magnetic fields. For example, spin polarization of electrons using non-polarized light through a chiral molecular structure such as DNA~\cite{gohler-dsDNA} has been studied. The measurements show that charge and spin transport are coupled, a phenomenon that is called chirality-induced spin selectivity (CISS) \cite{vanWeesCISS, EversCISS, RuitenbeekCISS}, which  chiral materials are abundant in biology (e.g. in sugars and proteins), however, as the complexity of these systems increases, so does the interpretation and understanding of such systems. In this sense, studies on simpler molecules are helpful to probe the effects of chirality on charge transport in further detail. In this regard, chiral structures such as helicenes have been proposed \cite{naamanHelicenes, matxain2019chirality, giaconi2023efficient, kettner2018chirality} due to their helical configuration and the possibility to isolate their mirror images, labeled as enantiomers.

At the single-molecule level, transport occurs out of equilibrium and is typically driven by an external bias voltage that induces a difference in the chemical potentials across the metallic leads. When a chiral molecule bridges the junction, the CISS effect states that moving electron spins with a particular orientation can distinguish chiral enantiomers, yielding a polarized current measurable even at room conditions. Consequently, it is essential to characterize charge transport in single molecule-metal junctions, which requires a basic understanding of their conductance and how the molecule connects. Additionally, theoretical calculations considering spin-orbit coupling aim to reproduce the behaviour, improving further the description of such systems~\cite{juanjoCISS, juanjoCISS2, CISSandNESpinAccumulation}, although discrepancies with transport experiments remain.

\begin{figure}[H]
\centering
\includegraphics[width=1\linewidth]{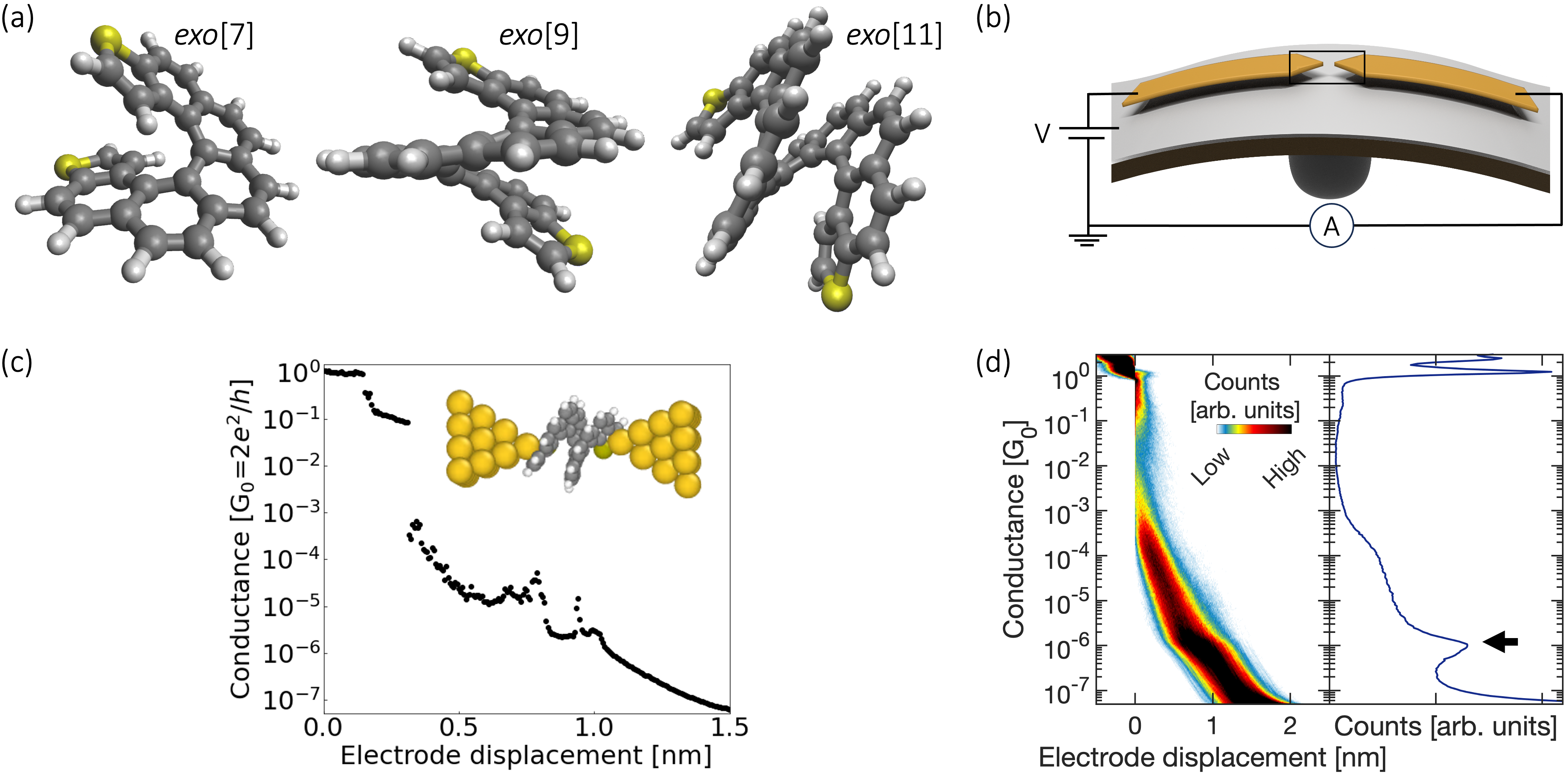}
\caption{MCBJ measurement methodology and molecular schemes. (a) Chemical structure of the dithia[n]helicenes: exo[7], exo[9] and exo[11]. Yellow, gray and white spheres represent sulfur, carbon and hydrogen atoms, respectively. (b) Schematic representation of the MCBJ set-up. Two gold electrodes with a notch are displayed. When the rod is pushed along the vertical direction, the junction stretches and a molecular conductor can be established. (c) Example of a breaking trace showing conductance in units of G$_0$ as a function of the relative displacement measured at a fixed bias voltage of $0.1~$V. Inset displays the exo[11]dithiahelicene candidate bridging between the gold electrodes. (d) 2D and 1D histograms of the raw data containing ten thousand consecutive breaking traces obtained for exo[11]dithiahelicene at 0.1~V. The black arrow points to the peak due to an amplifier artifact.}
\label{fig:scheme}
\end{figure}

Throughout this work, electron transport is investigated for a series of chiral molecules at ambient conditions using the mechanical controllable break junction (MCBJ) technique 
Our approach involves the study of set of helicenes with a helix structure and anchoring links incorporated to the molecular structure to enhance conductivity~\cite{[7]helicenes, [9]helicenes} (schematic illustrations are in Figure~\ref{fig:scheme}(a)). Henceforth, we focus on studying dithia[n]helicenes, where n represents the number of aromatic rings, specifically n = 7, 9, and 11. By analyzing these helicenes of varying sizes, we aim to gain insights of the nature of electronic transport mechanism in this system from the relationship between molecular length and electron transport characteristics.

In our molecules sulfur atoms are incorporated into the molecular structure using thiophene rings. In such a way, anchoring groups such  as thiols~\cite{thiolVenkataraman, Reedbenzenedithiol, thiolatedMolecules, Alkanedithiol, Zheng2018, cysteamine} are removed. This design choice originates from the idea of reducing potential barriers when the molecule bridges between electrodes while maintaining the conjugation throughout the molecule~\cite{theoConductancehelicenes}. This concept was supported by preliminary Density Functional Theory (DFT) calculations on isolated molecules, revealing the development of molecular orbitals that include the contribution of the sulfur atoms. Additionally, sulfur atoms are expected to provide mechanical stability while the helical structure offers its inherent flexibility~\cite{tuninghelicenes}. Furthermore, we explore the influence of the sulfur atoms by arranging the sulfurs in two different positions. First, the sulfurs located by facing outside the helical axis as illustrated in Figure~\ref{fig:scheme}(a) and second, the sulfurs positioned on the opposite side of the thiophene, facing the helical axis. We refer to these configurations as \textit{exo} and \textit{endo}, respectively. 

In our previous works \cite{[7]helicenes, [9]helicenes}, we described the preparation of both configurations for [7] and [9] dithiahelicene. The preparation of yet unreported exo[11] and endo[11] dithiahelicene was achieved by adapting these modular syntheses to a larger central phenanthrene fragment. This process involves our state of the art methods of Pd-catalyzed coupling reactions followed by a LED-driven final photocyclization step, as detailed in the Supplementary Information.

The MCBJ technique involves fixing the electrodes over a bending bead which provides high mechanical stability, as depicted in Figure~\ref{fig:scheme}(b). The electrodes consist of a gold junction that has been nanolithographed with a notch in the middle. A piezoelectric system applies a push to the junction which results in a controlled horizontal displacement, leading to elongation and finally the rupture at the notch. By cyclically bending and relaxing the electrodes, we create breaking and formation cycles, enabling precise control over the formation of metallic or molecular conductors. During this process, we measure the current flowing through the junction at a fixed bias voltage while stretching until it eventually breaks. 

Specifically, we record the evolution of the conductance (in terms of G$_0=2e^2/h$) as a function of the relative displacement between the leads forming a so-called breaking trace. 
An example of such a trace is depicted in Figure~\ref{fig:scheme}(c) while the rod is pushing until rupture. Additionally, we employ a logarithmic amplifier to achieve nine orders of magnitude during data acquisition~\cite{ornago2023}. By repeating the process thousands of times,  we generate 1D histograms to depict the distribution of conductance values obtained. Furthermore, by overlaying the breaking traces, we create 2D histograms that present a density plot of the evolution of most likely conductance values with separation of the electrodes. These 2D/1D histograms as depicted in Figure~\ref{fig:scheme}(d), provide valuable insights for further analysis. 
 
We have measured the electrical properties of the exo[7,9,11] and endo[11] molecules in dichloromethane (DCM) solution, considering both enantiomers. The measurements were performed with a concentration of 10 $\mu$M for exo[7] and 1 $\mu$M for the remaining molecules. For exo[7] we have recorded 2000 consecutive breaking traces while 10000 for the rest of the molecules. The use of low concentrations promotes single molecular bridges, although it may lead to a lower molecular yield. Figure~\ref{fig:scheme}(d) displays the collected data for the exo[11] helicenes in a 2D/1D histogram composition (as reference, Figure S3 of Supplementary Information depicts the 2D histograms of bare gold and the four molecules). Both histograms do not show clear peaks associated to single molecules because of the low molecular yield, which translates being the tunneling contribution the dominant one. An important observation in Figure \ref{fig:scheme}(d) is the presence of a peak around 10$^{-6}$ G$_0$, which is an artifact introduced by the logarithmic amplifier. This peak value remains consistent across all the collected data and shows the expected decrease in conductance value~\cite{ornago2023} as the bias voltage increases (see Supplementary Information, Figure S4, which showcases the evolution of the peak).

In order to extract meaningful features from the recorded datasets and filter out the influence of the logarithmic amplifier artifact, we performed a two-step analysis. Firstly, we utilized a neural network \cite{vanVeenNN2023} trained with a dataset comprising breaking traces of bare gold and with molecules. This model, incorporating dropout layers, allowed us to classify the breaking traces and identify those with molecules bridging the electrodes. Once the classification is completed, we employed the k-means++ clustering algorithm \cite{clustering}, an unsupervised machine learning technique, to subtract the underlying molecular information. This algorithm partitions the dataset into a predefined number of clusters. Through an iterative process, it identifies the most repeated values, so that traces with similar plateau structures converges into the same clusters. Consequently, this clustering analysis highlights the underlying molecular features within the predefined clusters. Details are specified in the Supplementary Information. Figure~S5 illustrates a composition of 2D/1D histograms of all ten clusters obtained for the exo[11] molecule measured at 0.1~V, including the tunneling ones.

\begin{figure}[H]
\centering
\includegraphics[width=1\linewidth]{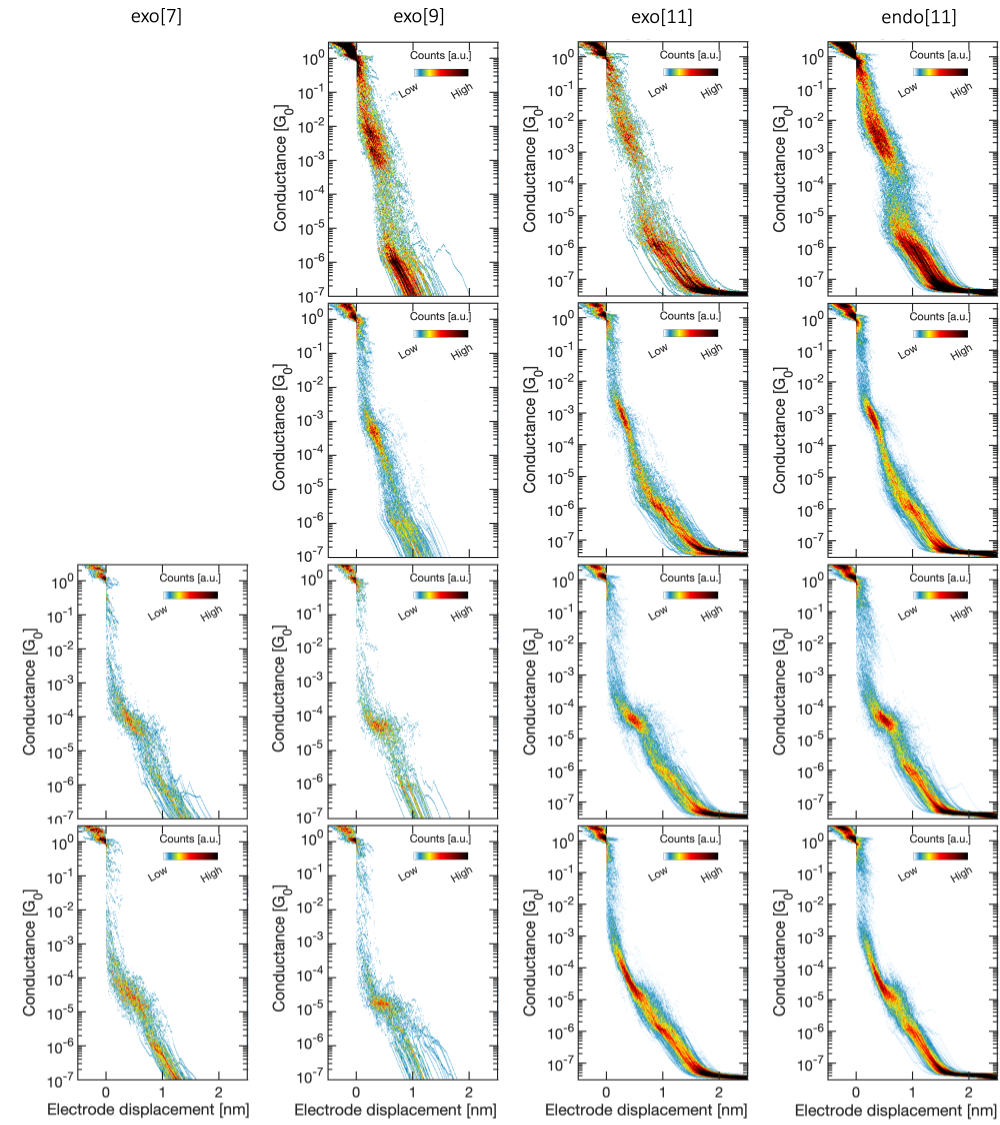}
\caption{2D histograms of the main clusters of each helical molecule at 0.1 V. Each column in the histogram corresponds to a specific molecule, showcasing different clusters associated with distinct plateau structures ranging from 10$^{-5}$ to 10$^{-3}$ G$_0$.}
\label{fig:2Dhistograms}
\end{figure}

Turning to the clusters that contain clear molecular features, Figure~\ref{fig:2Dhistograms} displays those for the data collected at 0.1 V for each molecule. Two main features can be obtained: First, the mean conductance value of the increased counts in the histogram (as derived from corresponding 1D histograms) and second, an estimated length of the molecular plateaus. To determine the mean conductance value ($\mu$) from the 1D histogram, we represent the logarithmic data in a logarithmic binning scale. Gaussian functions are then fitted to the data using the formula \begin{math}f(x) = \sum_{i=1}^{N}Ae^{{-({x-\mu})^2}/{2\sigma^2}}\end{math}. From this Gaussian fit the mean conductance value is then extracted (see Table S1 in Supplementary Information). In addition, we can define a lower conductance value within the cluster distribution as $\mu - 0.5 \sigma$. This value is used as criterion to limit the region of interest of the molecular plateaus. The length of each plateau is then computed with the start point fixed at the value of 0.3 G$_{0}$ and the end point being the lower conductance limit defined as $\mu - 0.5 \sigma$. After collecting the lengths of the plateaus, a 1D histogram is constructed. This histogram is then used to estimate the overall plateau length for each cluster through Gaussian fitting.

Figure \ref{fig:2Dhistograms} displays different clusters for each molecule taken at 0.1~V. The most prominent and frequently observed plateau structures converge into up to four assigned clusters. The mean conductance values falls within the range of 10$^{-3}$ to 10$^{-5}$ G$_0$. The first row displays a wider distribution, which may be indicative of at least two binding events. The exo[9] candidate exhibits a distribution that could contain two distinct peaks, however, this distinction is not consistently observed for the other molecules. Increasing the number of clusters when analyzing the dataset might provide a more detailed local view, but we are interested in capturing the global picture. Furthermore, using a higher parameter for clustering may result in overly segmented distributions.
To check for reproducibility and potential changes in conductance at higher voltage values~\cite{dependenceHOMOLUMO}, the measurements were repeated at different bias voltages. For exo[9, 11], the bias is ranged from 0.05 up to 0.35~V  and up to 1~V for endo[11]. To perform a statistical analysis, we recorded 10000 breaking traces, except for 0.7 and 1~V applied bias voltages, where 4000 and 2000 traces were obtained respectively, due to instabilities. The [9] and [11] helicenes were selected based on the idea that longer molecules could achieve a greater variety in binding configurations.  

\begin{figure}[H]
\centering
\includegraphics[width=1\linewidth]{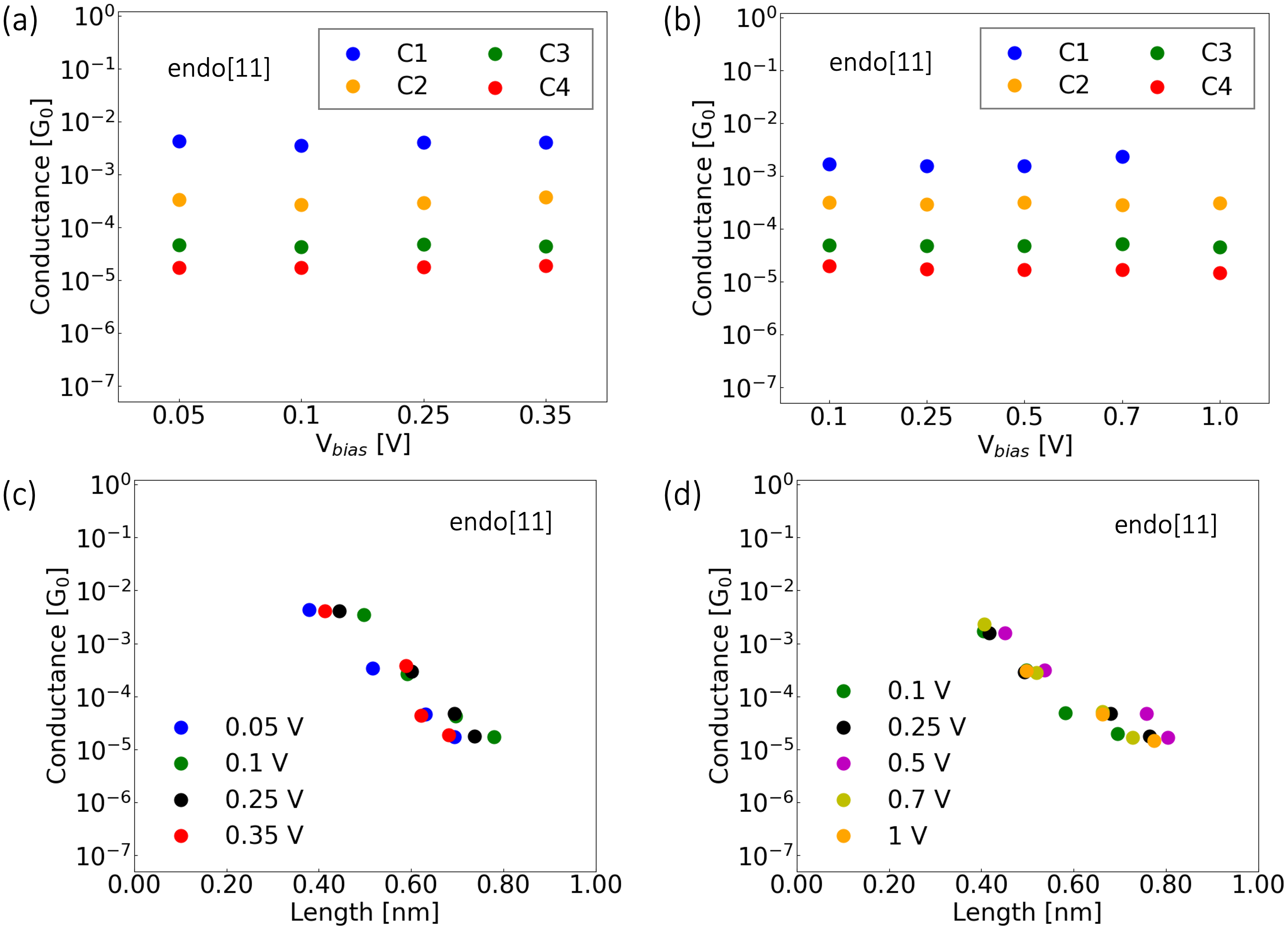}
\caption{Mean conductance values of the four clusters (C1-C4) found for two acquired data sets of endo[11] as a function of (a, b) the bias voltage and (c, d) the length. To improve clarity, common assignments in both (a) and (b) are color-coded identically, facilitating easy comparisons across various voltage levels. In contrast, panels (c, d) are color-coded based on the voltage dataset. Panels (a,c) refer to one data set analysed by using 10 clusters at low bias, while panels (b,d) represent another data set using 5 clusters for high bias voltages.}
\label{fig:followclusterSummary-endo11}
\end{figure}

The datasets measured at different bias voltages were simultaneously clustered, enabling the identification of common features as illustrated in Figures~S7 and S8 in the Supplementary Information. The fitted conductance values from each cluster measured  are presented in Figure~\ref{fig:followclusterSummary-endo11}(a, b) as a function of bias voltage. The figure shows that the same clusters appear across the different bias voltages and that the conductance values are largely insensitive to changes in bias.
This may indicate that the HOMO and LUMO are not close to the Fermi energy of the leads, i.e., the transmission function is nearly flat and far from the molecular orbitals. It is also noticeable that there is a slightly change from panel (a) to (b) toward lower conductance values, only for the data points shown in blue. Since we are clustering data from different bias voltages, including those up to 1 V, this last result may be indicative of the stability of molecular junctions at high bias voltages. 
Additionally, the information subtracted from the 2D/1D histograms enables the determination of the averaged length of the plateaus in each cluster; the mean conductance values are plotted against these lengths in Figure \ref{fig:followclusterSummary-endo11}(c,d). As observed, the blue data points exhibit similar lengths, despite showing slightly different conductance values. This may be a consequence of considering the same range of plateaus in both cases (panels (a) and (b)), indicating overall similar molecular events.
Among the studied molecules, a consistent trend emerges with a characteristic decay of conductance with average plateau length.  

Figure \ref{fig:followclusterSummary-endo11}(c) and (d) showcases a dependence with the length, meaning the possibility of contacting the molecules through different points and obtaining distinct conductance values. Furthermore, considering the repeatability and consistent trend observed, we calculated the average values of related clusters to determine the conductance value and the estimated length. One advantage of this approach is the large amount of data available for analysis. Additionally, when increasing the bias voltages, the molecular yield increases, possibly due to the electric field aiding the diffusion of molecules towards the tip. However, instabilities arise beyond 0.5 V, which poses a challenge for acquiring data. This drive us to compute the average for the available dataset ranging from 0.05 to 0.35 V. The conductance value for exo[7] helicene, without additional measurements, was obtained from the measurements at 0.1 V. The averaged conductance values are summarized in Table \ref{tab:conductance}.

\begin{table}[H]
\caption{Mean conductance values for all target molecules and different clusters. Conductance for exo[7] is obtained for $V_{\rm bias}=0.1$~V, while the mean values for the other molecules correspond to the average value among the different bias voltages up to 0.35~V. The clusters that show the most pronounced plateau structure are listed.}
\centering
\begin{tabular}{ccccc}
\hline
                               & exo{[}7{]}                                   & exo{[}9{]}                                   & exo{[}11{]}                                  & endo{[}11{]}                                 \\ \hline
G$_{\rm C1}$/G$_{0}$              &                                              & 2.0$\cdot$ 10$^{-3}$                         & 2.5 $\cdot$ 10$^{-3}$                        & 4.0 $\cdot$ 10$^{-3}$                        \\ \hline
G$_{\rm C2}$/G$_{0}$           &                                              & 1.7 $\cdot$ 10$^{-4}$ & 2.7 $\cdot$ 10$^{-3}$                        & 3.2 $\cdot$ 10$^{-3}$                        \\ \hline
G$_{\rm C3}$/G$_{0}$                               & 6.4 $\cdot$ 10$^{-5}$ & 3.2 $\cdot$ 10$^{-5}$                        & 5.8 $\cdot$ 10$^{-5}$ & 4.5 $\cdot$ 10$^{-5}$ \\ \hline
G$_{\rm C4}$/G$_{0}$ & 2.8 $\cdot$ 10$^{-5}$                        &                                              & 2.4 $\cdot$ 10$^{-5}$                        & 1.8 $\cdot$ 10$^{-5}$ \\ \hline
\end{tabular}
\label{tab:conductance}
\end{table}

\begin{figure}[H]
\centering
\includegraphics[width=0.85\linewidth]{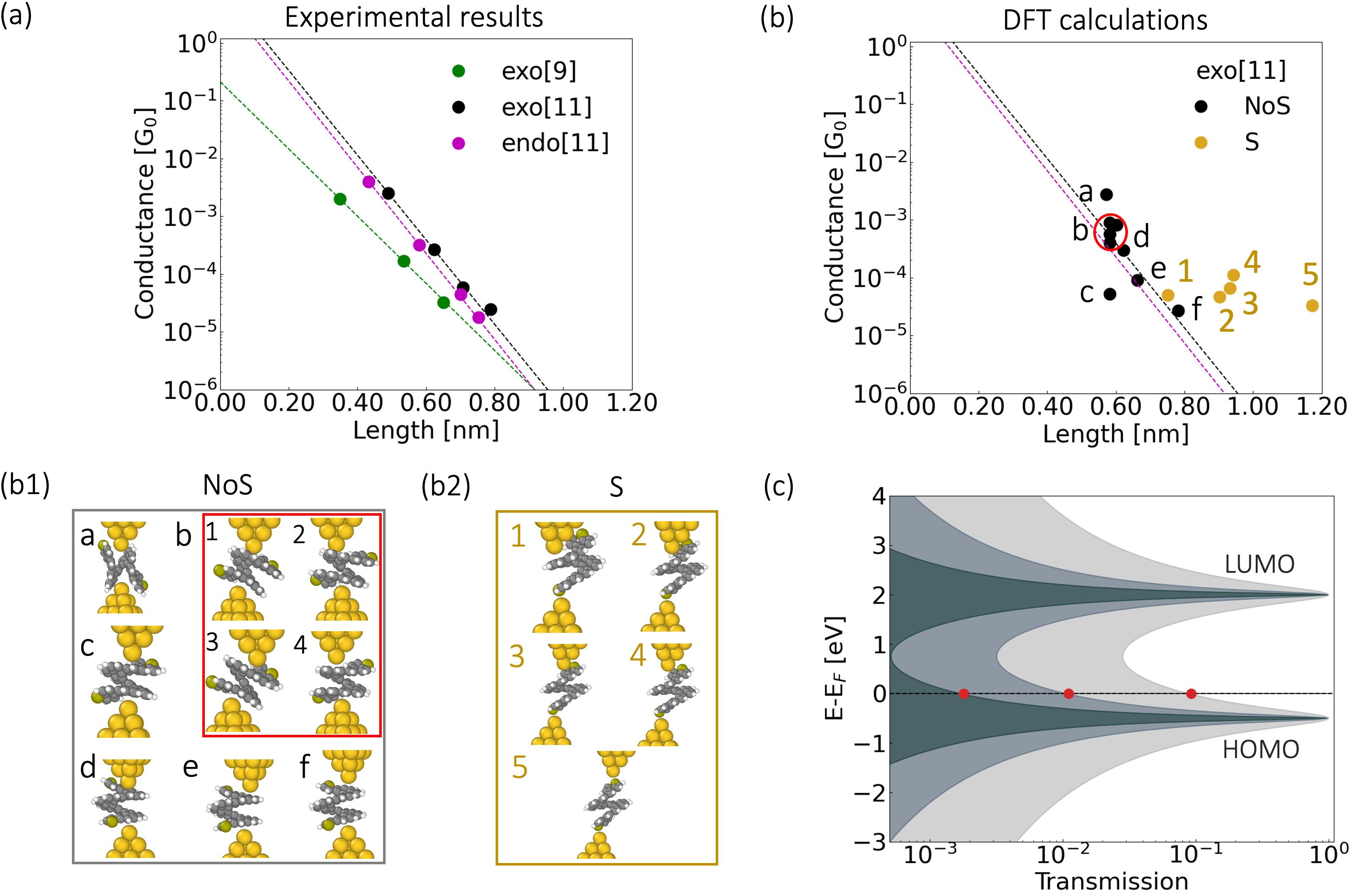}
\caption{Correlation between molecular structure and electronic conductance. (a) Averaged conductance as a function of the averaged length for the exo[9], endo[11] and exo[11] molecules. Dashed lines display the exponential fits. (b) Conductance data derived from DFT calculations, connected through the carbons, NoS, and through at least one of the sulfurs, S. Panels (b1) and (b2) depict the molecular structures associated with each data point in panel (b), illustrating the impact of molecular composition on conductance. (c) Schematic illustration depicting the off-resonant transport due to the energy misalignment between the molecular orbitals and the Fermi energy of the electrodes. Low conductance values are expected out of the resonance as the transport is mediated by the electrons at E=E$_F$, marked by the red dots.}
\label{fig:expFit}
\end{figure}

The averaged conductance vs. relative displacement for the longer helicenes is presented in Figure~\ref{fig:expFit}(a). Here, the conductance decays with the length of the molecular plateau, which seems to follow an exponential decrease. This trend can be fitted by following the expression \begin{math}G = G_{\rm c} e^{-\beta L}\end{math}, where $G$ is the conductance, the inverse of $G_{\rm c}$ defines the contact resistance with left and right anchors~\cite{contactresistance}, $\beta$ is the decay constant related to the tunnelling barrier and $L$ is the length. This exponential decay can be understood as a sign for off-resonant transport \cite{transportMechanism}, which significantly relies on the degree of localization and energy alignment of the molecular orbitals that mediate transport with the Fermi energy of the electrodes (sketched in panel (c) of Figure \ref{fig:expFit}). It is important to stress that this approach is commonly carried out by measuring the conductance for different molecules as a function of the number of repeating units in the molecule. In that case, by measuring the conductance of a series of fully stretched molecules of different sizes (molecular units), information on $\beta$ is obtained. Here, we use the approach to characterize charge transport along the molecule while it is stretched, as it will be contacted through different molecular positions, following the analysis on previous work on peptide chains~\cite{peptides}.

The $\beta$ parameter gives information on the distance of the Fermi energy of the electrodes to the molecular levels and whether the electrons responsible of the electronic transport through the molecule are more delocalized (small $\beta$ values) across the molecules or less localized (large $\beta$ values) on individual atoms. Thus, for similar alignments of the molecular and electrode energies, the systems can be considered to follow a trend typical of conjugated or non-conjugated compounds, respectively. 
Several studies have reported small decay values for $\beta$ related to conjugated systems such as oligophenylene and oligoacenes terminated in -S or -NC with $\beta$ 
around 0.050 nm$^{-1}$ or a molecular wire (DAD)$_n$ chain with $\beta=0.021$~nm$^{-1}$~\cite{molecularwire}. Smaller values for oligothiophenes were found for $\beta= 0.01$~nm$^{-1}$~\cite{Yamadaoligothiophenes}. On the other hand, there have been reported values bigger by an order of magnitude related to non-conjugated systems: alkanes with $\beta=7.5-10.0$ nm$^{-1}$~\cite{alkanes}, cysteamine-(n)glycinecysteine chains with $\beta=8.7$ nm$^{-1}$~\cite{cysteamine}, triglycine with $\beta=$ 9.7 nm$^{-1}$~\cite{triglycine} and peptide chains with $\beta'=13.5-15.3$~nm$^{-1}$~\cite{peptides}. In the latter case, not the plateau length but its mean position was considered, similar to what has been done in this paper. For each helicene, as depicted in Figure~\ref{fig:expFit}, an exponential fit was applied to obtain the $\beta$ parameter. The obtained $\beta$ values are $13.40  \pm  0.09$ and $16.9 \pm 0.2$ nm$^{-1}$ for the exo[9] and [11] and $17.18 \pm 0.08$ nm$^{-1}$ for the endo[11]. The differences in the $\beta$ parameter could be attributed to differences in the gap structure for both molecules.

We employed \textit{ab initio} calculations to support the experimental evidence and explore the influence of various molecular configurations on electronic transport. Density functional theory (DFT) was performed in order to optimize the geometries of the gold electrodes-molecule system, as displayed in Figure \ref{fig:expFit} (also refer to Figure S9 of the Supplementary Information for the transmission curves of the optimized scenarios). To compute the conductance, a combination of Spin-Orbit Coupling-corrected DFT with non-equilibrium Green’s function (NEGF) was employed using the ANT.GAUSSIAN code ~\cite{palacios2001fullerene, palacios2002transport, ANTG}. Moreover, the HSE06 functional was selected, which is extensively used in metal-organic systems \cite{HSEH1PBE03, HSEH1PBE04, camarasagómez2023transferable, JanusMATyHSE}. The Gaussian-type orbital basis sets employed in all electronic transport calculations in this manuscript are the same as those utilized by the authors in reference \cite{DednamACS2023}. In Figure \ref{fig:expFit}, panel (b) showcases the computed conductance values vs. the distance of the apex atoms of the gold electrodes, for all the optimized geometrical structures depicted in panels (b1) and (b2). The DFT conductances can be encompassed in two distinct situations: connection through the carbon atoms of the structure, noted as NoS, and connection through at least one sulfur atom, noted as S. Our first observation is regarding the impact of the molecular configurations on the conductance: connections through the helical structure (NoS) yield distinct signals which mainly depends on the distance and the geometry of the entire system, following what appears to be the observed experimental decay trends. However, once a sulfur atom participates in the transport, the conductance value seems to stabilize to values in the order of 10$^{-5}$ G$_0$, attributed to the barrier exert by the contact gold-sulfur.

Apparently, for the helicenes studied here, experimental results show predominantly up to four distinct conductance signatures, each associated with different anchor points on the molecule. Notably, there is a greater variety of contact configurations for longer molecules, leading to a common range of characteristic conductance plateaus for molecules of varying sizes. This scenario is supported by DFT calculations when the transport does not involve the sulfurs, which demonstrate that conductance is influenced by the binding geometry.

Although the connection between helicenes and the electrodes, as well as the number of molecular units involved, remains unknown, the decay ratios obtained fall within the same range as those reported in non-conjugated systems. This suggests that transport is carried out by the electrons out of the resonance, 
resulting in lower conductance. This decay is supported qualitatively by the DFT calculations as shown in Figure~\ref{fig:expFit}.

The consistent behavior of the mean conductance value, regardless of changes in bias voltage, suggests that the measurements are taken outside the resonant transport regime. This indicates that the molecular energy levels are misaligned with respect to the Fermi level of the electrodes, and transport occurs at some energy within the molecular energy gap. In addition, conductance length dependence measurements along each molecule, further suggest that off-resonant transport is likely the primary mechanism of transport for helicenes under ambient conditions. Our analysis supports the use of the $\beta$ decay analysis to get information on the details of the transport mechanism in molecular electronics. Furthermore, this result implies a limitation when detecting the CISS effect in these helicenes due to the low conductance values obtained and the misalignments of the energy levels of the system. This suggests the need to improve charge transport in these systems, which may involve modifying the molecular structure design or selecting different electrodes for conducting transport experiments. 


\begin{acknowledgement}
This work forms part of the Advanced Materials programme and was supported by MCIN with funding from European Union NextGenerationEU (PRTR-C17.I1) and by Generalitat Valenciana (MFA/2022/045). The authors acknowledge financial support from the Spanish Government through MFA/2022/045, PID2019-109539-GB-C41 and PID2022-141712NB-C22 and by the Generalitat Valenciana through PROMETEO/2021/017 and CIDEXG/2022/45. The theoretical modelling was performed on the high-performance computing facilities of the University of South Africa and the University of Alicante. C.H. and H.S.J.v.d.Z. acknowledge The Netherlands Organization for Scientifc Research (NWO; Natuurkunde Vrije Programma's: 680.90.18.01). 
\end{acknowledgement}

\bibliography{main.bib}

\providecommand{\latin}[1]{#1}
\makeatletter
\providecommand{\doi}
  {\begingroup\let\do\@makeother\dospecials
  \catcode`\{=1 \catcode`\}=2 \doi@aux}
\providecommand{\doi@aux}[1]{\endgroup\texttt{#1}}
\makeatother
\providecommand*\mcitethebibliography{\thebibliography}
\csname @ifundefined\endcsname{endmcitethebibliography}
  {\let\endmcitethebibliography\endthebibliography}{}
\begin{mcitethebibliography}{42}
\providecommand*\natexlab[1]{#1}
\providecommand*\mciteSetBstSublistMode[1]{}
\providecommand*\mciteSetBstMaxWidthForm[2]{}
\providecommand*\mciteBstWouldAddEndPuncttrue
  {\def\EndOfBibitem{\unskip.}}
\providecommand*\mciteBstWouldAddEndPunctfalse
  {\let\EndOfBibitem\relax}
\providecommand*\mciteSetBstMidEndSepPunct[3]{}
\providecommand*\mciteSetBstSublistLabelBeginEnd[3]{}
\providecommand*\EndOfBibitem{}
\mciteSetBstSublistMode{f}
\mciteSetBstMaxWidthForm{subitem}{(\alph{mcitesubitemcount})}
\mciteSetBstSublistLabelBeginEnd
  {\mcitemaxwidthsubitemform\space}
  {\relax}
  {\relax}

\bibitem[Zhang \latin{et~al.}(2015)Zhang, Zhong, Lin, Hu, Wu, Xu, Wee, and
  Chen]{singlemoleculeswitch}
Zhang,~J.~L.; Zhong,~J.~Q.; Lin,~J.~D.; Hu,~W.~P.; Wu,~K.; Xu,~G.~Q.; Wee,~A.
  T.~S.; Chen,~W. Towards single molecule switches. \emph{Chem. Soc. Rev.}
  \textbf{2015}, \emph{44}, 2998--3022\relax
\mciteBstWouldAddEndPuncttrue
\mciteSetBstMidEndSepPunct{\mcitedefaultmidpunct}
{\mcitedefaultendpunct}{\mcitedefaultseppunct}\relax
\EndOfBibitem
\bibitem[G{\"o}hler \latin{et~al.}(2011)G{\"o}hler, Hamelbeck, Markus, Kettner,
  Hanne, Vager, Naaman, and Zacharias]{gohler-dsDNA}
G{\"o}hler,~B.; Hamelbeck,~V.; Markus,~T.; Kettner,~M.; Hanne,~G.; Vager,~Z.;
  Naaman,~R.; Zacharias,~H. Spin selectivity in electron transmission through
  self-assembled monolayers of double-stranded DNA. \emph{Science}
  \textbf{2011}, \emph{331}, 894--897\relax
\mciteBstWouldAddEndPuncttrue
\mciteSetBstMidEndSepPunct{\mcitedefaultmidpunct}
{\mcitedefaultendpunct}{\mcitedefaultseppunct}\relax
\EndOfBibitem
\bibitem[Yang \latin{et~al.}(2020)Yang, van~der Wal, and van Wees]{vanWeesCISS}
Yang,~X.; van~der Wal,~C.~H.; van Wees,~B.~J. Detecting Chirality in
  Two-Terminal Electronic Nanodevices. \emph{Nano Letters} \textbf{2020},
  \emph{20}, 6148--6154, PMID: 32672980\relax
\mciteBstWouldAddEndPuncttrue
\mciteSetBstMidEndSepPunct{\mcitedefaultmidpunct}
{\mcitedefaultendpunct}{\mcitedefaultseppunct}\relax
\EndOfBibitem
\bibitem[Evers \latin{et~al.}(2022)Evers, Aharony, Bar‐Gill, Entin-Wohlman,
  Hedegård, Hod, Jelinek, Kamieniarz, Lemeshko, Michaeli, Mujica,
  Refaely-Abramson, Tal, Thijssen, Thoss, van Ruitenbeek, Venkataraman, and
  Kronik]{EversCISS}
Evers,~F. \latin{et~al.}  Theory of Chirality Induced Spin Selectivity:
  Progress and Challenges. \emph{Advanced Materials} \textbf{2022}, \emph{34},
  2106629\relax
\mciteBstWouldAddEndPuncttrue
\mciteSetBstMidEndSepPunct{\mcitedefaultmidpunct}
{\mcitedefaultendpunct}{\mcitedefaultseppunct}\relax
\EndOfBibitem
\bibitem[van Ruitenbeek \latin{et~al.}(2023)van Ruitenbeek, Korytár, and
  Evers]{RuitenbeekCISS}
van Ruitenbeek,~J.~M.; Korytár,~R.; Evers,~F. {Chirality-controlled spin
  scattering through quantum interference}. \emph{The Journal of Chemical
  Physics} \textbf{2023}, \emph{159}, 024710\relax
\mciteBstWouldAddEndPuncttrue
\mciteSetBstMidEndSepPunct{\mcitedefaultmidpunct}
{\mcitedefaultendpunct}{\mcitedefaultseppunct}\relax
\EndOfBibitem
\bibitem[Kiran \latin{et~al.}(2016)Kiran, Mathew, Cohen, Hern{\'a}ndez~Delgado,
  Lacour, and Naaman]{naamanHelicenes}
Kiran,~V.; Mathew,~S.~P.; Cohen,~S.~R.; Hern{\'a}ndez~Delgado,~I.; Lacour,~J.;
  Naaman,~R. Helicenes—A new class of organic spin filter. \emph{Advanced
  Materials} \textbf{2016}, \emph{28}, 1957--1962\relax
\mciteBstWouldAddEndPuncttrue
\mciteSetBstMidEndSepPunct{\mcitedefaultmidpunct}
{\mcitedefaultendpunct}{\mcitedefaultseppunct}\relax
\EndOfBibitem
\bibitem[Matxain \latin{et~al.}(2019)Matxain, Ugalde, Mujica, Allec, Wong, and
  Casanova]{matxain2019chirality}
Matxain,~J.~M.; Ugalde,~J.~M.; Mujica,~V.; Allec,~S.~I.; Wong,~B.~M.;
  Casanova,~D. Chirality Induced Spin Selectivity of Photoexcited Electrons in
  Carbon-Sulfur [n] Helicenes. \emph{ChemPhotoChem} \textbf{2019}, \emph{3},
  770--777\relax
\mciteBstWouldAddEndPuncttrue
\mciteSetBstMidEndSepPunct{\mcitedefaultmidpunct}
{\mcitedefaultendpunct}{\mcitedefaultseppunct}\relax
\EndOfBibitem
\bibitem[Giaconi \latin{et~al.}(2023)Giaconi, Poggini, Lupi, Briganti, Kumar,
  Das, Sorrentino, Viglianisi, Menichetti, Naaman, \latin{et~al.}
  others]{giaconi2023efficient}
Giaconi,~N.; Poggini,~L.; Lupi,~M.; Briganti,~M.; Kumar,~A.; Das,~T.~K.;
  Sorrentino,~A.~L.; Viglianisi,~C.; Menichetti,~S.; Naaman,~R., \latin{et~al.}
   Efficient Spin-Selective Electron Transport at Low Voltages of Thia-Bridged
  Triarylamine Hetero [4] helicenes Chemisorbed Monolayer. \emph{ACS nano}
  \textbf{2023}, \emph{17}, 15189--15198\relax
\mciteBstWouldAddEndPuncttrue
\mciteSetBstMidEndSepPunct{\mcitedefaultmidpunct}
{\mcitedefaultendpunct}{\mcitedefaultseppunct}\relax
\EndOfBibitem
\bibitem[Kettner \latin{et~al.}(2018)Kettner, Maslyuk, N{\"u}renberg, Seibel,
  Gutierrez, Cuniberti, Ernst, and Zacharias]{kettner2018chirality}
Kettner,~M.; Maslyuk,~V.~V.; N{\"u}renberg,~D.; Seibel,~J.; Gutierrez,~R.;
  Cuniberti,~G.; Ernst,~K.-H.; Zacharias,~H. Chirality-dependent electron spin
  filtering by molecular monolayers of helicenes. \emph{The journal of physical
  chemistry letters} \textbf{2018}, \emph{9}, 2025--2030\relax
\mciteBstWouldAddEndPuncttrue
\mciteSetBstMidEndSepPunct{\mcitedefaultmidpunct}
{\mcitedefaultendpunct}{\mcitedefaultseppunct}\relax
\EndOfBibitem
\bibitem[Dednam \latin{et~al.}(2023)Dednam, García-Blázquez, Zotti, Lombardi,
  Sabater, Pakdel, and Palacios]{juanjoCISS}
Dednam,~W.; García-Blázquez,~M.~A.; Zotti,~L.~A.; Lombardi,~E.~B.;
  Sabater,~C.; Pakdel,~S.; Palacios,~J.~J. A Group-Theoretic Approach to the
  Origin of Chirality-Induced Spin-Selectivity in Nonmagnetic Molecular
  Junctions. \emph{ACS Nano} \textbf{2023}, \emph{17}, 6452--6465, PMID:
  36947721\relax
\mciteBstWouldAddEndPuncttrue
\mciteSetBstMidEndSepPunct{\mcitedefaultmidpunct}
{\mcitedefaultendpunct}{\mcitedefaultseppunct}\relax
\EndOfBibitem
\bibitem[García-Blázquez \latin{et~al.}(2023)García-Blázquez, Dednam, and
  Palacios]{juanjoCISS2}
García-Blázquez,~M.~A.; Dednam,~W.; Palacios,~J.~J. Nonequilibrium
  Magneto-Conductance as a Manifestation of Spin Filtering in Chiral
  Nanojunctions. \emph{The Journal of Physical Chemistry Letters}
  \textbf{2023}, \emph{14}, 7931--7939, PMID: 37646507\relax
\mciteBstWouldAddEndPuncttrue
\mciteSetBstMidEndSepPunct{\mcitedefaultmidpunct}
{\mcitedefaultendpunct}{\mcitedefaultseppunct}\relax
\EndOfBibitem
\bibitem[Naskar \latin{et~al.}(2023)Naskar, Mujica, and
  Herrmann]{CISSandNESpinAccumulation}
Naskar,~S.; Mujica,~V.; Herrmann,~C. Chiral-Induced Spin Selectivity and
  Non-equilibrium Spin Accumulation in Molecules and Interfaces: A
  First-Principles Study. \emph{The Journal of Physical Chemistry Letters}
  \textbf{2023}, \emph{14}, 694--701, PMID: 36638217\relax
\mciteBstWouldAddEndPuncttrue
\mciteSetBstMidEndSepPunct{\mcitedefaultmidpunct}
{\mcitedefaultendpunct}{\mcitedefaultseppunct}\relax
\EndOfBibitem
\bibitem[Baciu \latin{et~al.}(2020)Baciu, de~Ara, Sabater, Untiedt, and
  Guijarro]{[7]helicenes}
Baciu,~B.~C.; de~Ara,~T.; Sabater,~C.; Untiedt,~C.; Guijarro,~A. Helical
  nanostructures for organic electronics: the role of topological sulfur in ad
  hoc synthesized dithia [7] helicenes studied in the solid state and on a gold
  surface. \emph{Nanoscale Advances} \textbf{2020}, \emph{2}, 1921--1926\relax
\mciteBstWouldAddEndPuncttrue
\mciteSetBstMidEndSepPunct{\mcitedefaultmidpunct}
{\mcitedefaultendpunct}{\mcitedefaultseppunct}\relax
\EndOfBibitem
\bibitem[Baciu \latin{et~al.}(2022)Baciu, Bronk, de~Ara, Rodriguez, Morgante,
  Vanthuyne, Sabater, Untiedt, Autschbach, Crassous, and
  Guijarro]{[9]helicenes}
Baciu,~B.~C.; Bronk,~P.~J.; de~Ara,~T.; Rodriguez,~R.; Morgante,~P.;
  Vanthuyne,~N.; Sabater,~C.; Untiedt,~C.; Autschbach,~J.; Crassous,~J.;
  Guijarro,~A. Dithia[9]helicenes: Molecular design{,} surface imaging{,} and
  circularly polarized luminescence with enhanced dissymmetry factors. \emph{J.
  Mater. Chem. C} \textbf{2022}, \emph{10}, 14306--14318\relax
\mciteBstWouldAddEndPuncttrue
\mciteSetBstMidEndSepPunct{\mcitedefaultmidpunct}
{\mcitedefaultendpunct}{\mcitedefaultseppunct}\relax
\EndOfBibitem
\bibitem[Venkataraman \latin{et~al.}(2006)Venkataraman, Klare, Tam, Nuckolls,
  Hybertsen, and Steigerwald]{thiolVenkataraman}
Venkataraman,~L.; Klare,~J.; Tam,~I.; Nuckolls,~C.; Hybertsen,~M.;
  Steigerwald,~M. Single-Molecule Circuits with Well-Defined Molecular
  Conductance. \emph{Nano letters} \textbf{2006}, \emph{6}, 458--62\relax
\mciteBstWouldAddEndPuncttrue
\mciteSetBstMidEndSepPunct{\mcitedefaultmidpunct}
{\mcitedefaultendpunct}{\mcitedefaultseppunct}\relax
\EndOfBibitem
\bibitem[Reed \latin{et~al.}(1997)Reed, Zhou, Muller, Burgin, and
  Tour]{Reedbenzenedithiol}
Reed,~M.~A.; Zhou,~C.; Muller,~C.~J.; Burgin,~T.~P.; Tour,~J.~M. Conductance of
  a Molecular Junction. \emph{Science} \textbf{1997}, \emph{278},
  252--254\relax
\mciteBstWouldAddEndPuncttrue
\mciteSetBstMidEndSepPunct{\mcitedefaultmidpunct}
{\mcitedefaultendpunct}{\mcitedefaultseppunct}\relax
\EndOfBibitem
\bibitem[Ulrich \latin{et~al.}(2006)Ulrich, Esrail, Pontius, Venkataraman,
  Millar, and Doerrer]{thiolatedMolecules}
Ulrich,~J.; Esrail,~D.; Pontius,~W.; Venkataraman,~L.; Millar,~D.;
  Doerrer,~L.~H. Variability of Conductance in Molecular Junctions. \emph{The
  Journal of Physical Chemistry B} \textbf{2006}, \emph{110}, 2462--2466, PMID:
  16471840\relax
\mciteBstWouldAddEndPuncttrue
\mciteSetBstMidEndSepPunct{\mcitedefaultmidpunct}
{\mcitedefaultendpunct}{\mcitedefaultseppunct}\relax
\EndOfBibitem
\bibitem[Li \latin{et~al.}(2008)Li, Pobelov, Wandlowski, Bagrets, Arnold, and
  Evers]{Alkanedithiol}
Li,~C.; Pobelov,~I.; Wandlowski,~T.; Bagrets,~A.; Arnold,~A.; Evers,~F. Charge
  Transport in Single Au | Alkanedithiol | Au Junctions:â€‰ Coordination
  Geometries and Conformational Degrees of Freedom. \emph{Journal of the
  American Chemical Society} \textbf{2008}, \emph{130}, 318--326, PMID:
  18076172\relax
\mciteBstWouldAddEndPuncttrue
\mciteSetBstMidEndSepPunct{\mcitedefaultmidpunct}
{\mcitedefaultendpunct}{\mcitedefaultseppunct}\relax
\EndOfBibitem
\bibitem[Zheng \latin{et~al.}(2018)Zheng, Liu, Zhuo, Li, Jin, Yang, Chen, Shi,
  Xiao, Hong, and Tian]{Zheng2018}
Zheng,~J.; Liu,~J.; Zhuo,~Y.; Li,~R.; Jin,~X.; Yang,~Y.; Chen,~Z.-B.; Shi,~J.;
  Xiao,~Z.; Hong,~W.; Tian,~Z.-q. Electrical and SERS detection of
  disulfide-mediated dimerization in single-molecule benzene-1{,}4-dithiol
  junctions. \emph{Chem. Sci.} \textbf{2018}, \emph{9}, 5033--5038\relax
\mciteBstWouldAddEndPuncttrue
\mciteSetBstMidEndSepPunct{\mcitedefaultmidpunct}
{\mcitedefaultendpunct}{\mcitedefaultseppunct}\relax
\EndOfBibitem
\bibitem[Xiao \latin{et~al.}(2004)Xiao, Xu, and Tao]{cysteamine}
Xiao,; Xu,; Tao, Conductance Titration of Single-Peptide Molecules.
  \emph{Journal of the American Chemical Society} \textbf{2004}, \emph{126},
  5370--5371, PMID: 15113203\relax
\mciteBstWouldAddEndPuncttrue
\mciteSetBstMidEndSepPunct{\mcitedefaultmidpunct}
{\mcitedefaultendpunct}{\mcitedefaultseppunct}\relax
\EndOfBibitem
\bibitem[Treboux \latin{et~al.}(1999)Treboux, Lapstun, Wu, and
  Silverbrook]{theoConductancehelicenes}
Treboux,~G.; Lapstun,~P.; Wu,~Z.; Silverbrook,~K. Electronic conductance of
  helicenes. \emph{Chemical Physics Letters} \textbf{1999}, \emph{301},
  493--497\relax
\mciteBstWouldAddEndPuncttrue
\mciteSetBstMidEndSepPunct{\mcitedefaultmidpunct}
{\mcitedefaultendpunct}{\mcitedefaultseppunct}\relax
\EndOfBibitem
\bibitem[Vacek \latin{et~al.}(2015)Vacek, Chocholoušová, Stará, Starý, and
  Dubi]{tuninghelicenes}
Vacek,~J.; Chocholoušová,~J.~V.; Stará,~I.~G.; Starý,~I.; Dubi,~Y.
  Mechanical tuning of conductance and thermopower in helicene molecular
  junctions. \emph{Nanoscale} \textbf{2015}, \emph{7}, 8793--8802\relax
\mciteBstWouldAddEndPuncttrue
\mciteSetBstMidEndSepPunct{\mcitedefaultmidpunct}
{\mcitedefaultendpunct}{\mcitedefaultseppunct}\relax
\EndOfBibitem
\bibitem[Ornago(2023)]{ornago2023}
Ornago,~L. Complexity of Electron Transport in Nanoscale Molecular Junctions.
  Doctoral thesis, Delft University of Technology, 2023; TU Delft QN/van der
  Zant Lab\relax
\mciteBstWouldAddEndPuncttrue
\mciteSetBstMidEndSepPunct{\mcitedefaultmidpunct}
{\mcitedefaultendpunct}{\mcitedefaultseppunct}\relax
\EndOfBibitem
\bibitem[van Veen \latin{et~al.}(2023)van Veen, Ornago, van~der Zant, and
  El~Abbassi]{vanVeenNN2023}
van Veen,~F.; Ornago,~L.; van~der Zant,~H.~S.; El~Abbassi,~M. Generalized
  neural network approach for separation of molecular breaking traces.
  \emph{Journal of Materials Chemistry C} \textbf{2023}, \relax
\mciteBstWouldAddEndPunctfalse
\mciteSetBstMidEndSepPunct{\mcitedefaultmidpunct}
{}{\mcitedefaultseppunct}\relax
\EndOfBibitem
\bibitem[Cabosart \latin{et~al.}(2019)Cabosart, Abbassi, Stefani, Frisenda,
  Calame, Zant, and Perrin]{clustering}
Cabosart,~D.; Abbassi,~M.; Stefani,~D.; Frisenda,~R.; Calame,~M.; Zant,~H.;
  Perrin,~M. A reference-free clustering method for the analysis of molecular
  break-junction measurements. \emph{Applied Physics Letters} \textbf{2019},
  \emph{114}, 143102\relax
\mciteBstWouldAddEndPuncttrue
\mciteSetBstMidEndSepPunct{\mcitedefaultmidpunct}
{\mcitedefaultendpunct}{\mcitedefaultseppunct}\relax
\EndOfBibitem
\bibitem[Koch \latin{et~al.}(2012)Koch, Ample, Joachim, and
  Grill]{dependenceHOMOLUMO}
Koch,~M.; Ample,~F.; Joachim,~C.; Grill,~L. Voltage-dependent conductance of a
  single graphene nanoribbon. \emph{Nature nanotechnology} \textbf{2012},
  \emph{7}\relax
\mciteBstWouldAddEndPuncttrue
\mciteSetBstMidEndSepPunct{\mcitedefaultmidpunct}
{\mcitedefaultendpunct}{\mcitedefaultseppunct}\relax
\EndOfBibitem
\bibitem[Engelkes \latin{et~al.}(2004)Engelkes, Beebe, and
  Frisbie]{contactresistance}
Engelkes,~V.~B.; Beebe,~J.~M.; Frisbie,~C.~D. Length-Dependent Transport in
  Molecular Junctions Based on SAMs of Alkanethiols and Alkanedithiols: Effect
  of Metal Work Function and Applied Bias on Tunneling Efficiency and Contact
  Resistance. \emph{Journal of the American Chemical Society} \textbf{2004},
  \emph{126}, 14287--14296, PMID: 15506797\relax
\mciteBstWouldAddEndPuncttrue
\mciteSetBstMidEndSepPunct{\mcitedefaultmidpunct}
{\mcitedefaultendpunct}{\mcitedefaultseppunct}\relax
\EndOfBibitem
\bibitem[Moth-Poulsen and Bj{\o}rnholm(2009)Moth-Poulsen, and
  Bj{\o}rnholm]{transportMechanism}
Moth-Poulsen,~K.; Bj{\o}rnholm,~T. Molecular electronics with single molecules
  in solid-state devices. \emph{Nature nanotechnology} \textbf{2009}, \emph{4},
  551--556\relax
\mciteBstWouldAddEndPuncttrue
\mciteSetBstMidEndSepPunct{\mcitedefaultmidpunct}
{\mcitedefaultendpunct}{\mcitedefaultseppunct}\relax
\EndOfBibitem
\bibitem[Stefani \latin{et~al.}(2021)Stefani, Guo, Ornago, Cabosart,
  El~Abbassi, Sheves, Cahen, and van~der Zant]{peptides}
Stefani,~D.; Guo,~C.; Ornago,~L.; Cabosart,~D.; El~Abbassi,~M.; Sheves,~M.;
  Cahen,~D.; van~der Zant,~H. S.~J. Conformation-dependent charge transport
  through short peptides. \emph{Nanoscale} \textbf{2021}, \emph{13},
  3002--3009\relax
\mciteBstWouldAddEndPuncttrue
\mciteSetBstMidEndSepPunct{\mcitedefaultmidpunct}
{\mcitedefaultendpunct}{\mcitedefaultseppunct}\relax
\EndOfBibitem
\bibitem[Nacci \latin{et~al.}(2015)Nacci, Ample, Bleger, Hecht, Joachim, and
  Grill]{molecularwire}
Nacci,~C.; Ample,~F.; Bleger,~D.; Hecht,~S.; Joachim,~C.; Grill,~L. Conductance
  of a single flexible molecular wire composed of alternating donor and
  acceptor units. \emph{Nature communications} \textbf{2015}, \emph{6},
  7397\relax
\mciteBstWouldAddEndPuncttrue
\mciteSetBstMidEndSepPunct{\mcitedefaultmidpunct}
{\mcitedefaultendpunct}{\mcitedefaultseppunct}\relax
\EndOfBibitem
\bibitem[Yamada \latin{et~al.}(2008)Yamada, Kumazawa, Noutoshi, Tanaka, and
  Tada]{Yamadaoligothiophenes}
Yamada,~R.; Kumazawa,~H.; Noutoshi,~T.; Tanaka,~S.; Tada,~H. Electrical
  Conductance of Oligothiophene Molecular Wires. \emph{Nano Letters}
  \textbf{2008}, \emph{8}, 1237--1240, PMID: 18311936\relax
\mciteBstWouldAddEndPuncttrue
\mciteSetBstMidEndSepPunct{\mcitedefaultmidpunct}
{\mcitedefaultendpunct}{\mcitedefaultseppunct}\relax
\EndOfBibitem
\bibitem[van Veen \latin{et~al.}(2022)van Veen, Ornago, van~der Zant, and
  El~Abbassi]{alkanes}
van Veen,~F.~H.; Ornago,~L.; van~der Zant,~H. S.~J.; El~Abbassi,~M. Benchmark
  Study of Alkane Molecular Chains. \emph{The Journal of Physical Chemistry C}
  \textbf{2022}, \emph{126}, 8801--8806\relax
\mciteBstWouldAddEndPuncttrue
\mciteSetBstMidEndSepPunct{\mcitedefaultmidpunct}
{\mcitedefaultendpunct}{\mcitedefaultseppunct}\relax
\EndOfBibitem
\bibitem[Park \latin{et~al.}(2007)Park, Whalley, Kamenetska, Steigerwald,
  Hybertsen, Nuckolls, and Venkataraman]{triglycine}
Park,~Y.~S.; Whalley,~A.~C.; Kamenetska,~M.; Steigerwald,~M.~L.;
  Hybertsen,~M.~S.; Nuckolls,~C.; Venkataraman,~L. Contact Chemistry and
  Single-Molecule Conductance: A Comparison of Phosphines, Methyl Sulfides, and
  Amines. \emph{Journal of the American Chemical Society} \textbf{2007},
  \emph{129}, 15768--15769, PMID: 18052282\relax
\mciteBstWouldAddEndPuncttrue
\mciteSetBstMidEndSepPunct{\mcitedefaultmidpunct}
{\mcitedefaultendpunct}{\mcitedefaultseppunct}\relax
\EndOfBibitem
\bibitem[Palacios \latin{et~al.}(2001)Palacios, P\'erez-Jim\'enez, Louis, and
  Verg\'es]{palacios2001fullerene}
Palacios,~J.~J.; P\'erez-Jim\'enez,~A.~J.; Louis,~E.; Verg\'es,~J.~A.
  Fullerene-based molecular nanobridges: A first-principles study. \emph{Phys.
  Rev. B} \textbf{2001}, \emph{64}, 115411\relax
\mciteBstWouldAddEndPuncttrue
\mciteSetBstMidEndSepPunct{\mcitedefaultmidpunct}
{\mcitedefaultendpunct}{\mcitedefaultseppunct}\relax
\EndOfBibitem
\bibitem[Palacios \latin{et~al.}(2002)Palacios, P\'erez-Jim\'enez, Louis,
  SanFabi{\'a}n, and Verg\'es]{palacios2002transport}
Palacios,~J.~J.; P\'erez-Jim\'enez,~A.~J.; Louis,~E.; SanFabi{\'a}n,~E.;
  Verg\'es,~J.~A. First-principles approach to electrical transport in
  atomic-scale nanostructures. \emph{Phys. Rev. B} \textbf{2002}, \emph{66},
  035322\relax
\mciteBstWouldAddEndPuncttrue
\mciteSetBstMidEndSepPunct{\mcitedefaultmidpunct}
{\mcitedefaultendpunct}{\mcitedefaultseppunct}\relax
\EndOfBibitem
\bibitem[Dednam \latin{et~al.}()Dednam, Zotti, and Palacios]{ANTG}
Dednam,~W.; Zotti,~L.~A.; Palacios,~J.~J. Computer code {ANT.G}aussian, with
  {SOC} corrections. Available from
  \url{https://github.com/juanjosepalacios/ANT.Gaussian}, Date of access:
  15-Feb-2023\relax
\mciteBstWouldAddEndPuncttrue
\mciteSetBstMidEndSepPunct{\mcitedefaultmidpunct}
{\mcitedefaultendpunct}{\mcitedefaultseppunct}\relax
\EndOfBibitem
\bibitem[Heyd \latin{et~al.}(2003)Heyd, Scuseria, and Ernzerhof]{HSEH1PBE03}
Heyd,~J.; Scuseria,~G.~E.; Ernzerhof,~M. {Hybrid functionals based on a
  screened Coulomb potential}. \emph{The Journal of Chemical Physics}
  \textbf{2003}, \emph{118}, 8207--8215\relax
\mciteBstWouldAddEndPuncttrue
\mciteSetBstMidEndSepPunct{\mcitedefaultmidpunct}
{\mcitedefaultendpunct}{\mcitedefaultseppunct}\relax
\EndOfBibitem
\bibitem[Heyd and Scuseria(2004)Heyd, and Scuseria]{HSEH1PBE04}
Heyd,~J.; Scuseria,~G.~E. {Efficient hybrid density functional calculations in
  solids: Assessment of the Heyd–Scuseria–Ernzerhof screened Coulomb hybrid
  functional}. \emph{The Journal of Chemical Physics} \textbf{2004},
  \emph{121}, 1187--1192\relax
\mciteBstWouldAddEndPuncttrue
\mciteSetBstMidEndSepPunct{\mcitedefaultmidpunct}
{\mcitedefaultendpunct}{\mcitedefaultseppunct}\relax
\EndOfBibitem
\bibitem[Camarasa-Gómez \latin{et~al.}(2023)Camarasa-Gómez, Ramasubramaniam,
  Neaton, and Kronik]{camarasagómez2023transferable}
Camarasa-Gómez,~M.; Ramasubramaniam,~A.; Neaton,~J.~B.; Kronik,~L.
  Transferable screened range-separated hybrid functionals for electronic and
  optical properties of van der Waals materials. 2023\relax
\mciteBstWouldAddEndPuncttrue
\mciteSetBstMidEndSepPunct{\mcitedefaultmidpunct}
{\mcitedefaultendpunct}{\mcitedefaultseppunct}\relax
\EndOfBibitem
\bibitem[Yin \latin{et~al.}(2021)Yin, Tan, Ding, Wen, Li, Teobaldi, and
  Liu]{JanusMATyHSE}
Yin,~W.-J.; Tan,~H.-J.; Ding,~P.-J.; Wen,~B.; Li,~X.-B.; Teobaldi,~G.;
  Liu,~L.-M. Recent advances in low-dimensional Janus materials: theoretical
  and simulation perspectives. \emph{Mater. Adv.} \textbf{2021}, \emph{2},
  7543--7558\relax
\mciteBstWouldAddEndPuncttrue
\mciteSetBstMidEndSepPunct{\mcitedefaultmidpunct}
{\mcitedefaultendpunct}{\mcitedefaultseppunct}\relax
\EndOfBibitem
\bibitem[Dednam \latin{et~al.}(2023)Dednam, García-Blázquez, Zotti, Lombardi,
  Sabater, Pakdel, and Palacios]{DednamACS2023}
Dednam,~W.; García-Blázquez,~M.~A.; Zotti,~L.~A.; Lombardi,~E.~B.;
  Sabater,~C.; Pakdel,~S.; Palacios,~J.~J. A Group-Theoretic Approach to the
  Origin of Chirality-Induced Spin-Selectivity in Nonmagnetic Molecular
  Junctions. \emph{ACS Nano} \textbf{2023}, \emph{17}, 6452--6465, PMID:
  36947721\relax
\mciteBstWouldAddEndPuncttrue
\mciteSetBstMidEndSepPunct{\mcitedefaultmidpunct}
{\mcitedefaultendpunct}{\mcitedefaultseppunct}\relax
\EndOfBibitem
\end{mcitethebibliography}


\providecommand{\latin}[1]{#1}
\makeatletter
\providecommand{\doi}
  {\begingroup\let\do\@makeother\dospecials
  \catcode`\{=1 \catcode`\}=2 \doi@aux}
\providecommand{\doi@aux}[1]{\endgroup\texttt{#1}}
\makeatother
\providecommand*\mcitethebibliography{\thebibliography}
\csname @ifundefined\endcsname{endmcitethebibliography}
  {\let\endmcitethebibliography\endthebibliography}{}
\begin{mcitethebibliography}{11}
\providecommand*\natexlab[1]{#1}
\providecommand*\mciteSetBstSublistMode[1]{}
\providecommand*\mciteSetBstMaxWidthForm[2]{}
\providecommand*\mciteBstWouldAddEndPuncttrue
  {\def\EndOfBibitem{\unskip.}}
\providecommand*\mciteBstWouldAddEndPunctfalse
  {\let\EndOfBibitem\relax}
\providecommand*\mciteSetBstMidEndSepPunct[3]{}
\providecommand*\mciteSetBstSublistLabelBeginEnd[3]{}
\providecommand*\EndOfBibitem{}
\mciteSetBstSublistMode{f}
\mciteSetBstMaxWidthForm{subitem}{(\alph{mcitesubitemcount})}
\mciteSetBstSublistLabelBeginEnd
  {\mcitemaxwidthsubitemform\space}
  {\relax}
  {\relax}

\bibitem[Baciu \latin{et~al.}(2020)Baciu, de~Ara, Sabater, Untiedt, and
  Guijarro]{[7]helicenes}
Baciu,~B.~C.; de~Ara,~T.; Sabater,~C.; Untiedt,~C.; Guijarro,~A. Helical
  nanostructures for organic electronics: the role of topological sulfur in ad
  hoc synthesized dithia [7] helicenes studied in the solid state and on a gold
  surface. \emph{Nanoscale Advances} \textbf{2020}, \emph{2}, 1921--1926\relax
\mciteBstWouldAddEndPuncttrue
\mciteSetBstMidEndSepPunct{\mcitedefaultmidpunct}
{\mcitedefaultendpunct}{\mcitedefaultseppunct}\relax
\EndOfBibitem
\bibitem[Baciu \latin{et~al.}(2022)Baciu, Bronk, de~Ara, Rodriguez, Morgante,
  Vanthuyne, Sabater, Untiedt, Autschbach, Crassous, and
  Guijarro]{[9]helicenes}
Baciu,~B.~C.; Bronk,~P.~J.; de~Ara,~T.; Rodriguez,~R.; Morgante,~P.;
  Vanthuyne,~N.; Sabater,~C.; Untiedt,~C.; Autschbach,~J.; Crassous,~J.
  \latin{et~al.}  Dithia[9]helicenes: Molecular design{,} surface imaging{,}
  and circularly polarized luminescence with enhanced dissymmetry factors.
  \emph{J. Mater. Chem. C} \textbf{2022}, \emph{10}, 14306--14318\relax
\mciteBstWouldAddEndPuncttrue
\mciteSetBstMidEndSepPunct{\mcitedefaultmidpunct}
{\mcitedefaultendpunct}{\mcitedefaultseppunct}\relax
\EndOfBibitem
\bibitem[Heyd \latin{et~al.}(2003)Heyd, Scuseria, and Ernzerhof]{HSEH1PBE03}
Heyd,~J.; Scuseria,~G.~E.; Ernzerhof,~M. {Hybrid functionals based on a
  screened Coulomb potential}. \emph{The Journal of Chemical Physics}
  \textbf{2003}, \emph{118}, 8207--8215\relax
\mciteBstWouldAddEndPuncttrue
\mciteSetBstMidEndSepPunct{\mcitedefaultmidpunct}
{\mcitedefaultendpunct}{\mcitedefaultseppunct}\relax
\EndOfBibitem
\bibitem[Heyd and Scuseria(2004)Heyd, and Scuseria]{HSEH1PBE04}
Heyd,~J.; Scuseria,~G.~E. {Efficient hybrid density functional calculations in
  solids: Assessment of the Heyd–Scuseria–Ernzerhof screened Coulomb hybrid
  functional}. \emph{The Journal of Chemical Physics} \textbf{2004},
  \emph{121}, 1187--1192\relax
\mciteBstWouldAddEndPuncttrue
\mciteSetBstMidEndSepPunct{\mcitedefaultmidpunct}
{\mcitedefaultendpunct}{\mcitedefaultseppunct}\relax
\EndOfBibitem
\bibitem[Camarasa-Gómez \latin{et~al.}(2023)Camarasa-Gómez, Ramasubramaniam,
  Neaton, and Kronik]{camarasagómez2023transferable}
Camarasa-Gómez,~M.; Ramasubramaniam,~A.; Neaton,~J.~B.; Kronik,~L.
  Transferable screened range-separated hybrid functionals for electronic and
  optical properties of van der Waals materials. 2023\relax
\mciteBstWouldAddEndPuncttrue
\mciteSetBstMidEndSepPunct{\mcitedefaultmidpunct}
{\mcitedefaultendpunct}{\mcitedefaultseppunct}\relax
\EndOfBibitem
\bibitem[Yin \latin{et~al.}(2021)Yin, Tan, Ding, Wen, Li, Teobaldi, and
  Liu]{JanusMATyHSE}
Yin,~W.-J.; Tan,~H.-J.; Ding,~P.-J.; Wen,~B.; Li,~X.-B.; Teobaldi,~G.;
  Liu,~L.-M. Recent advances in low-dimensional Janus materials: theoretical
  and simulation perspectives. \emph{Mater. Adv.} \textbf{2021}, \emph{2},
  7543--7558\relax
\mciteBstWouldAddEndPuncttrue
\mciteSetBstMidEndSepPunct{\mcitedefaultmidpunct}
{\mcitedefaultendpunct}{\mcitedefaultseppunct}\relax
\EndOfBibitem
\bibitem[Palacios \latin{et~al.}(2001)Palacios, P\'erez-Jim\'enez, Louis, and
  Verg\'es]{palacios2001fullerene}
Palacios,~J.~J.; P\'erez-Jim\'enez,~A.~J.; Louis,~E.; Verg\'es,~J.~A.
  Fullerene-based molecular nanobridges: A first-principles study. \emph{Phys.
  Rev. B} \textbf{2001}, \emph{64}, 115411\relax
\mciteBstWouldAddEndPuncttrue
\mciteSetBstMidEndSepPunct{\mcitedefaultmidpunct}
{\mcitedefaultendpunct}{\mcitedefaultseppunct}\relax
\EndOfBibitem
\bibitem[Palacios \latin{et~al.}(2002)Palacios, P\'erez-Jim\'enez, Louis,
  SanFabi{\'a}n, and Verg\'es]{palacios2002transport}
Palacios,~J.~J.; P\'erez-Jim\'enez,~A.~J.; Louis,~E.; SanFabi{\'a}n,~E.;
  Verg\'es,~J.~A. First-principles approach to electrical transport in
  atomic-scale nanostructures. \emph{Phys. Rev. B} \textbf{2002}, \emph{66},
  035322\relax
\mciteBstWouldAddEndPuncttrue
\mciteSetBstMidEndSepPunct{\mcitedefaultmidpunct}
{\mcitedefaultendpunct}{\mcitedefaultseppunct}\relax
\EndOfBibitem
\bibitem[Dednam \latin{et~al.}()Dednam, Zotti, and Palacios]{ANTG}
Dednam,~W.; Zotti,~L.~A.; Palacios,~J.~J. Computer code {ANT.G}aussian, with
  {SOC} corrections. Available from
  \url{https://github.com/juanjosepalacios/ANT.Gaussian}, Date of access:
  15-Feb-2023\relax
\mciteBstWouldAddEndPuncttrue
\mciteSetBstMidEndSepPunct{\mcitedefaultmidpunct}
{\mcitedefaultendpunct}{\mcitedefaultseppunct}\relax
\EndOfBibitem
\bibitem[Dednam \latin{et~al.}(2023)Dednam, García-Blázquez, Zotti, Lombardi,
  Sabater, Pakdel, and Palacios]{DednamACS2023}
Dednam,~W.; García-Blázquez,~M.~A.; Zotti,~L.~A.; Lombardi,~E.~B.;
  Sabater,~C.; Pakdel,~S.; Palacios,~J.~J. A Group-Theoretic Approach to the
  Origin of Chirality-Induced Spin-Selectivity in Nonmagnetic Molecular
  Junctions. \emph{ACS Nano} \textbf{2023}, \emph{17}, 6452--6465, PMID:
  36947721\relax
\mciteBstWouldAddEndPuncttrue
\mciteSetBstMidEndSepPunct{\mcitedefaultmidpunct}
{\mcitedefaultendpunct}{\mcitedefaultseppunct}\relax
\EndOfBibitem
\end{mcitethebibliography}

\end{document}


\section{Supplementary Information}

\subsection{Synthesis of dithia[n]helicenes: exo[7], exo[9], exo[11] and endo[11] dithiahelicenes}

Fragments \textbf{1}, \textbf{2} and \textbf{3} in the synthetic scheme below are described in our previous works \cite{[7]helicenes, [9]helicenes}. The details and characterization of bis-stilbenic precursors \textbf{4} and \textbf{5} and final dithiahelicenes exo[11] (\textbf{6}) and endo[11] (\textbf{7}) follows.

Bis-stilbenic intermediates \textbf{4} and \textbf{5} needed to obtain the final dithia[11]helicenes were prepared by a Suzuki-Miyaura coupling between 3,6-dibromophenanthrene (\textbf{3}) as the central fragment \cite{[7]helicenes}, and the corresponding isomer of vinyl boronic pinacol esters (\textbf{1} or \textbf{2}) \cite{[9]helicenes}. These boronic esters bear the exo-/endo- structural features in the type of fusion of their terminal thiophene ring, and so is eventually transferred to the final helical compound. Yields of products are given in parenthesis (Figure  S1). 

\begin{figure}[H]
\centering
\renewcommand\thefigure{S1}  
\includegraphics[width=1\linewidth]{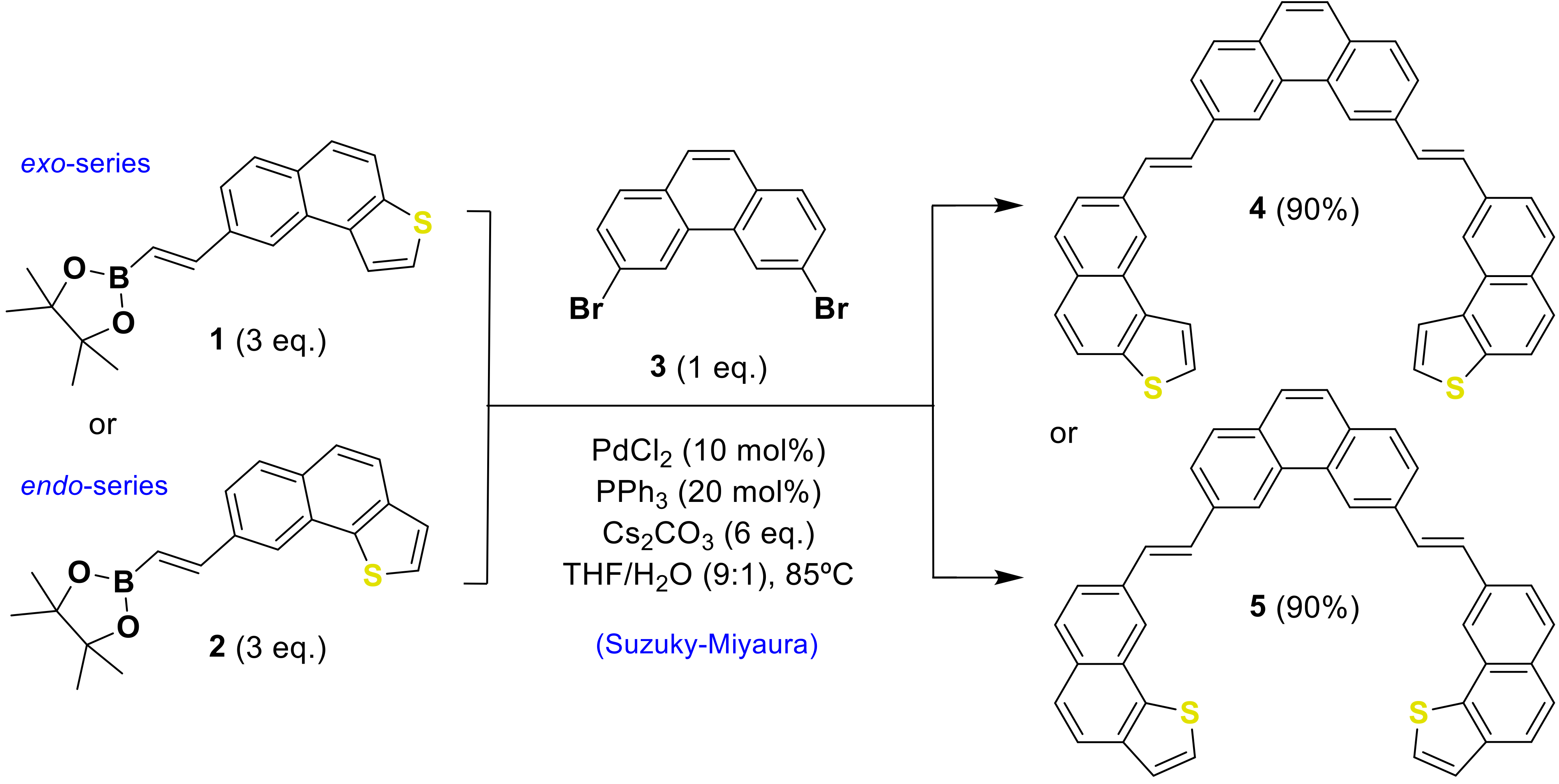}
\caption{Synthesis of bis-stilbenic precursors 4 and 5 for exo- and endo-dithia[11]helicenes}
\end{figure}

Compounds \textbf{4} and \textbf{5} are the photochemical precursors of the new dithia[11]helicenes, namely exo[11] and endo[11], respectively. The key photochemical step is a double Mallory reaction. It was performed employing a 365 nm light source from LED plates, using iodine as oxidant and an excess of 1,2-butyleneoxide as an acid quencher. The reactions were performed under inert atmosphere (Ar) (Figure S2). The characterization of all new compounds, i.e. both bis-stilbenic precursors and dithia[11]helicenes is described in the Supporting Information.

\begin{figure}[H]
\centering
\renewcommand\thefigure{S2}  
\includegraphics[width=1\linewidth]{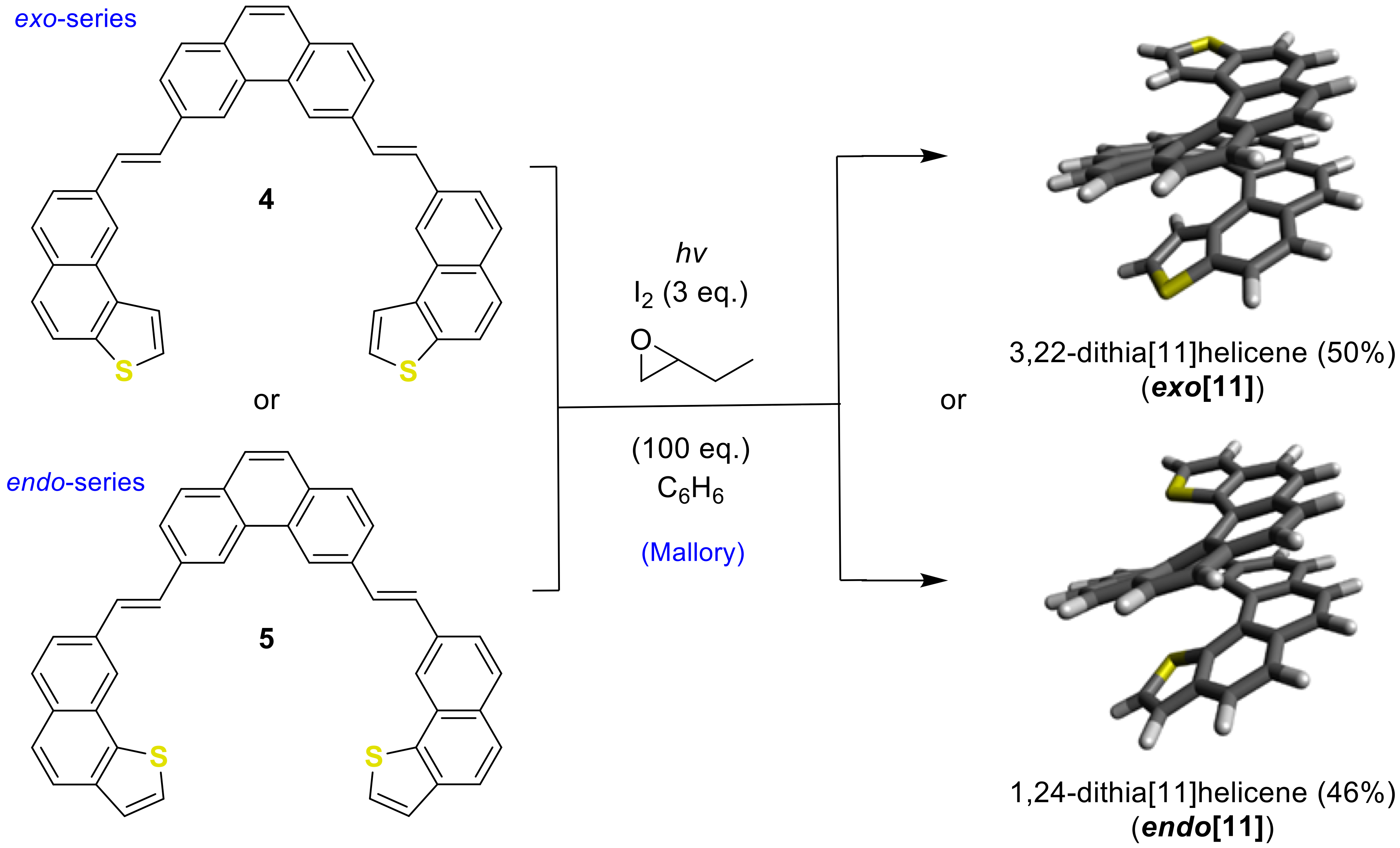}
\caption{Final oxidative photocyclization step obtaining exo- and endo-dithia[11]helicenes.}
\end{figure}

\subsection{Synthesis and characterization of bis-stilbenic precursors 4 and 5, exo-dithia[11]helicene (6) and endo-dithia[11]helicene (7)}

\subsection{3,6-bis((E)-2-(naphtho[2,1-b]tiophen-8-yl)vinyl)phenanthrene (4)}

The synthesis of this compound was carried out by a Suzuki-Miyaura reaction under the following conditions: In an over-dried pressure tube PdCl$_2$ (4.4 mg; 0.025 mmol; 0.10 eq.), PPh$_3$ (13.1 mg; 0.05 mmol; 0.20 eq.), Cs$_2$CO$_3$ (488.73 mg; 1.5 mmol; 6 eq.), 3,6-dibromophenanthrene (84.10 mg; 0.25 mmol; 1 eq.) were added. The tube was sealed with a septum and after three cycles of vacuum/argon, (E)-4,4,5,5-tetramethyl-2-(2-(naphtho[2,1-b]tiophen-8-yl)vinyl)-1,3,2-dioxoborolane (\textbf{1}) (252.11 mg, 0.75 mmol, 3 eq.) dissolved in 1.8 mL of THF were added, followed by 0.2 mL of H$_2$O, both of them via syringe. After that, the tube was closed and heated in an oil bath at 85℃ for 20 hours. An insoluble solid in suspension was observed in the tube. This insoluble yellow solid was filtered, washed with H$_2$O and CH$_2$Cl$_2$ and used without further purification. 

\begin{figure}[H]
\centering
\includegraphics[width=.3\linewidth]{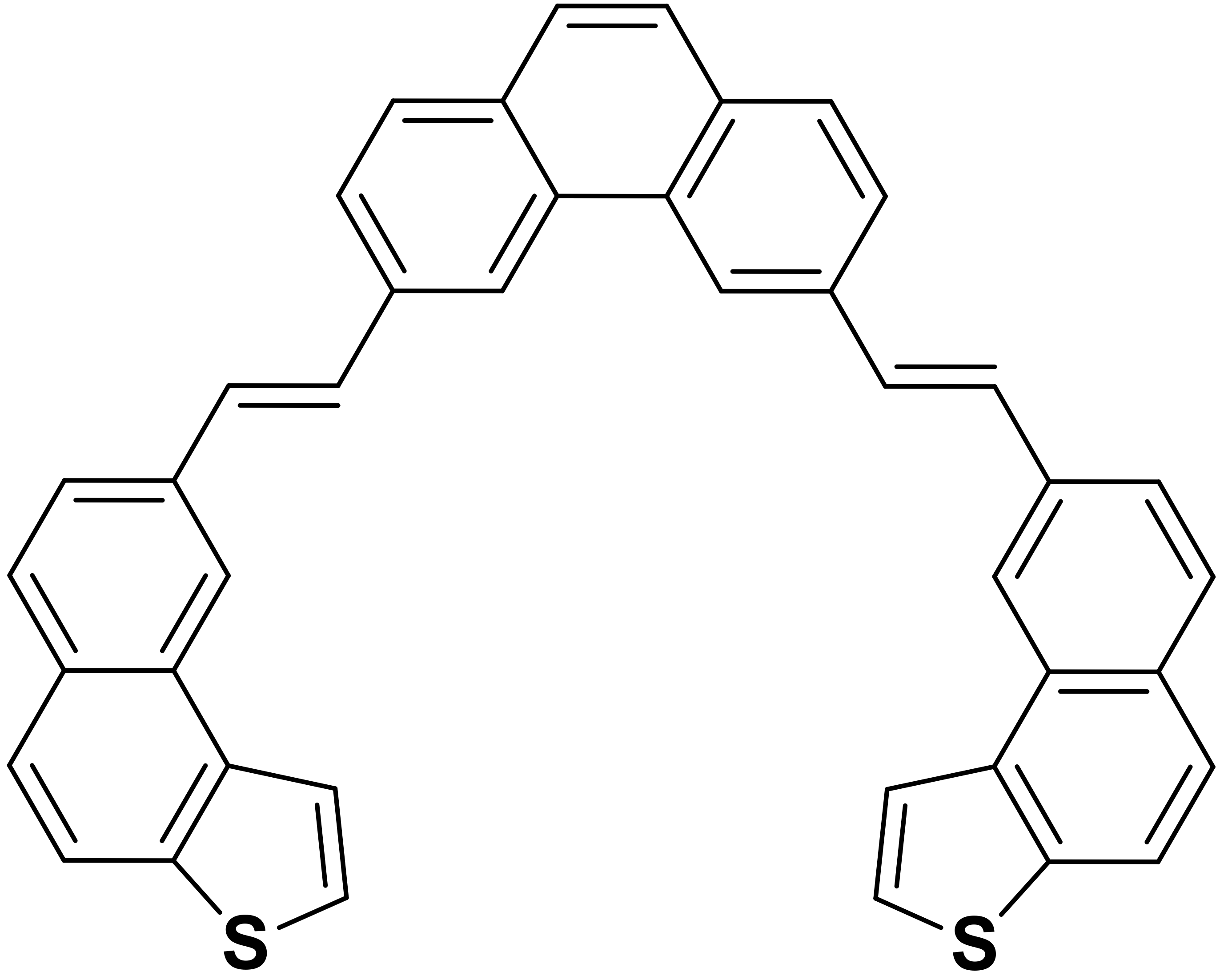}
\end{figure}

Yellow solid in 90\% yield; MS (EI, DIP) m/z: 596.1 (M$^+$+2, 20), 595.1 (M$^+$+1, 48), 594.1 (M$^{+}$, 100), 545.1 (28), 544.1 (71), 408.1 (13), 358.1 (14), 297.1 (20), 272.0 (15), 197.0 (11). IR (neat) $\nu_{max}$: 3062, 3020, 1616, 1373, 1192, 1153, 953, 876, 833, 710, 663 cm$^{-1}$.

\subsection{3,6-bis((E)-2-(naphtho[1,2-b]tiophen-8-yl)vinyl)phenanthrene (5)}

This compound was prepared following the previous procedure, but using (E)-4,4,5,5-tetramethyl-2-(2-(naphtho[1,2-b]tiophen-8-yl)vinyl)-1,3,2-dioxoborolane (\textbf{2}) as starting reagent.

\begin{figure}[H]
\centering
\includegraphics[width=.3\linewidth]{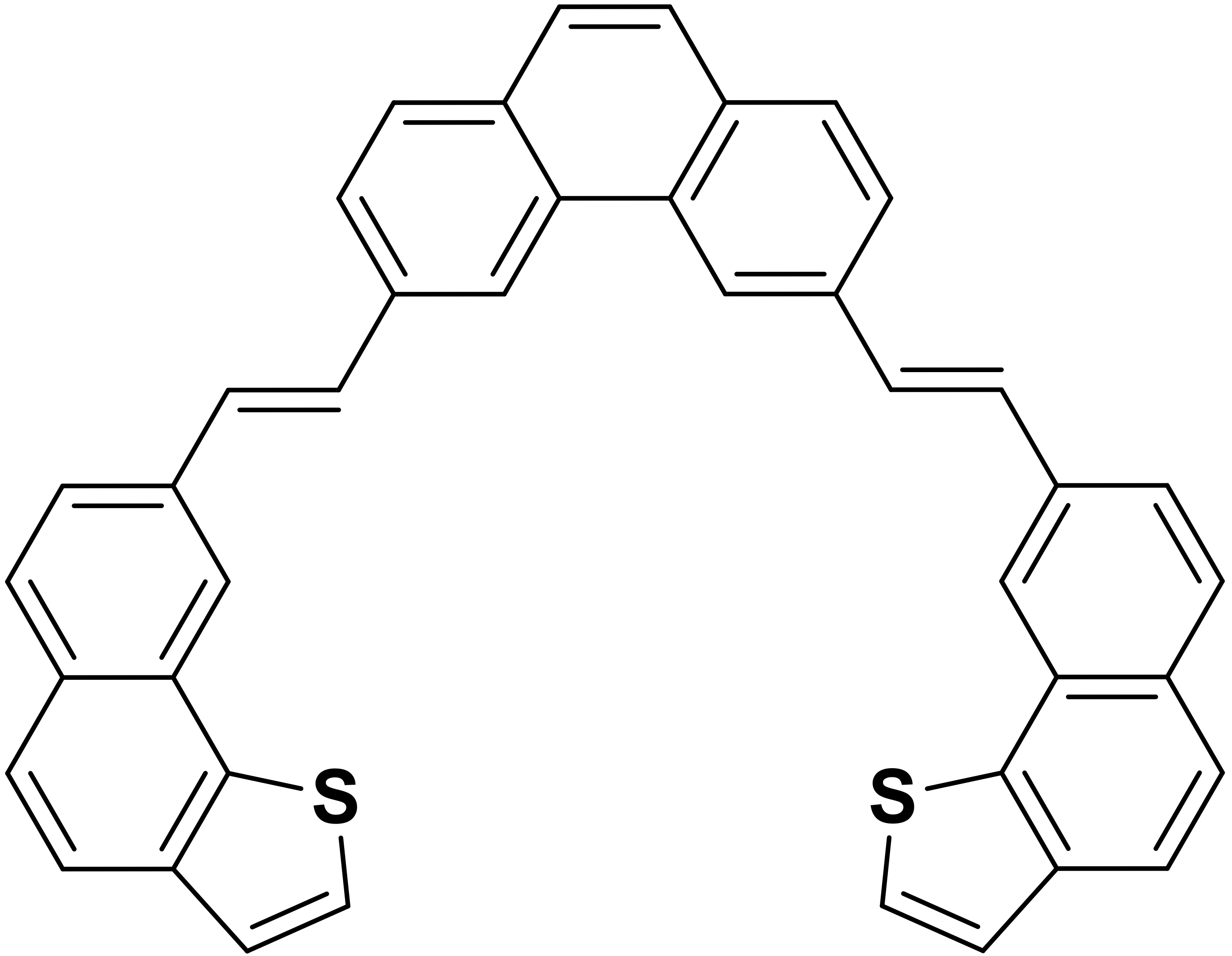}
\end{figure}

Yellow solid in 90\% yield; MS (EI, DIP) m/z 596.1 (M$^+$+2, 20), 595.1 (M$^+$+1, 47), 594.1 (M$^+$, 100), 546.1 (16), 545.1 (30), 544.1 (70), 408.1 (13), 384.1 (8), 358.1 (13), 297.1 (26), 272.0 (17), 197.0 (14). IR (neat) $\nu_{max}$: 2974, 2885, 1612, 1311, 1261, 1088, 1045, 949, 876, 833, 702 cm$^{-1}$.

\subsection{3,22-dithia[11]helicene (exo-dithia[11]helicene) (6)}
In an oven-dried 250mL Schlenk tube 3,6-bis((E)-2-(naphtho[2,1-b]tiophen-8-yl)vinyl)phenanthrene (\textbf{4}) (17.84 mg, 0.03 mmol, 1 eq.) was added, followed by 200 mL of benzene. The mixture was stirred and heated a little bit to dissolve the reagent. After that, iodine (22.84 mg, 0.09 mmol, 3 eq.)  and 1,2-butyleneoxide (0.3 mL, 3 mmol, 100 eq.) were added. The solution was bubbled with Ar for 15 minutes and then two 265 nm LEDs plates of 50 W each were turned on. With the light on, the solution was bubbled with Ar for 15 more minutes and then the tube was closed. The mixture was irradiated with LEDs overnight. After the reaction was completed, it was washed with aqueous NaHSO$_3$, then H$_2$O, dried with anhydrous magnesium sulphate, filtered and the solvent evaporated under reduced pressure (15 Torr). The residue was purified by column chromatography on silica gel (hexane-CH$_2$Cl$_2$ 8:2) to obtain a yellow solid. The product was recrystallized in CH$_2$Cl$_2$ to obtain yellow crystals.

\begin{figure}[H]
\centering
\includegraphics[width=.2\linewidth]{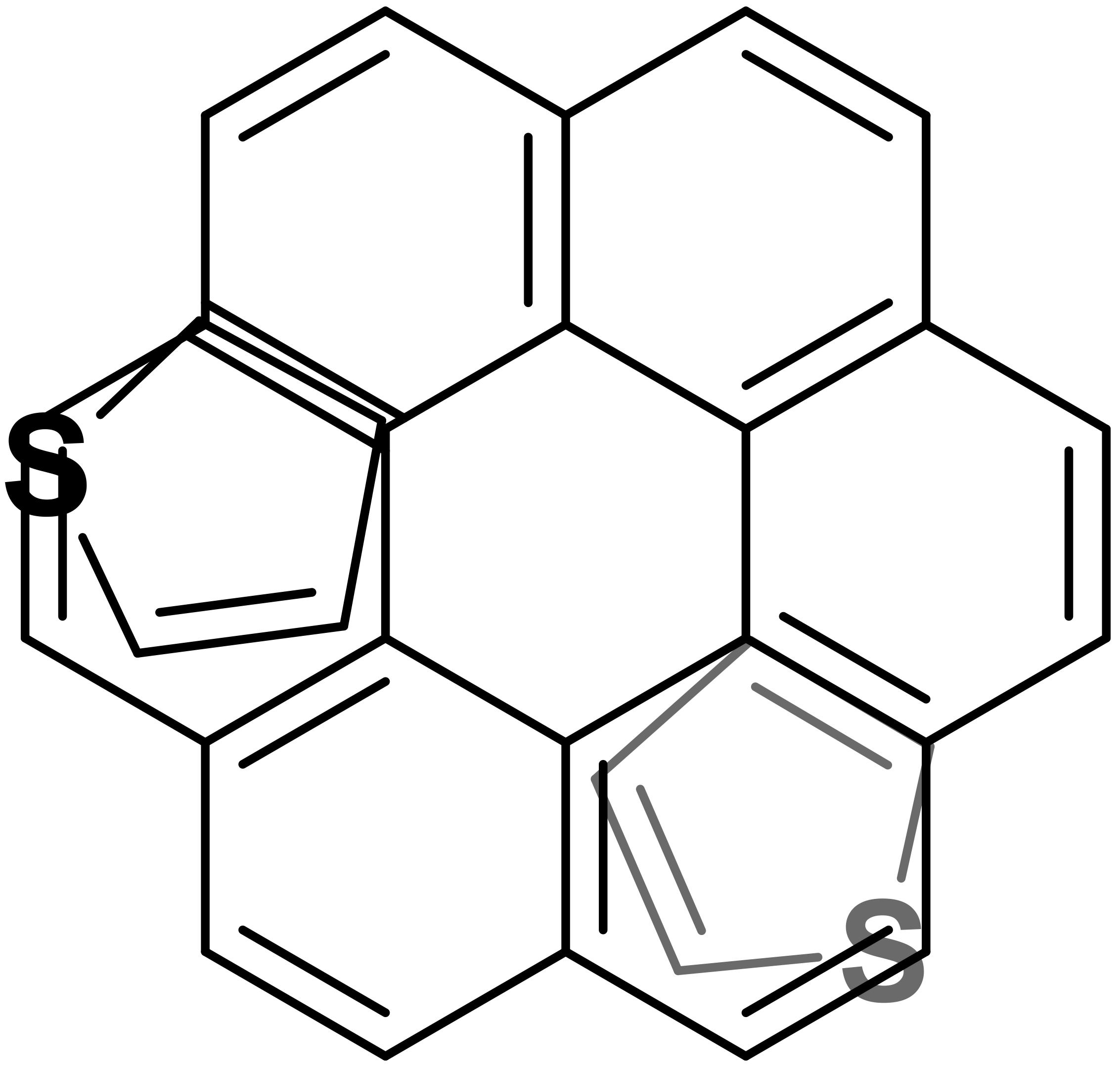}
\end{figure}

Yellow crystals in 50\% yield; R$_f$= 0.45 (hexane-CH$_2$Cl$_2$ 6:4); $^1$H-NMR (CDCl$_3$, 400MHz): $\delta$= 7.61 (s, 2H), 7.47 (dd, J= 8.5, 0.8 Hz, 2H), 7.38 (d, J= 8.2 Hz, 2H), 7.34 (d, J= 8.4 Hz, 2H), 7.27 (d, J= 8.3 Hz, 2H), 7.23 (d, J= 8.3 Hz, 2H), 7.22 (d, J= 8.2 Hz, 2H), 7.22 (s, 4H), 6.37 (dd, J= 5.6, 0.4 Hz, 2H), 5.81 (dd, J= 5.6, 0.8 Hz, 2H) ppm. $^{13}$C-NMR (CDCl$_3$, 101 MHz): $\delta$=137.44 (-C-, 2C), 133.97 (-C-, 2C), 131.95 (-C-, 2C), 131.10 (-CH-, 2C), 130.04 (-C-, 2C), 127.24 (-C-, 2C), 126.76 (-CH-, 2C), 126.46 (-CH-, 2H), 126.18 (-CH-, 2C), 126.14 (-CH-, 2C), 125.63 (-CH-, 2C), 125.54 (-CH-, 2C), 124.62 (-CH-, 2C), 124.57 (-C-, 2C), 124.46 (-C-, 2C), 123.51 (-C-, 2C), 122.72 (-CH-, 2C), 121.59 (-CH-, 2C), 121.08 (-C-, 2C), 119.85 (-CH-, 2C), 118.91 (-C-, 2C) ppm. MS (EI, DIP) m/z: 592.2 (M$^+$+2, 20), 591.2 (M$^+$+1, 49), 590.2 (M$^+$, 100), 306.1 (12), 295.1 (26), 282.1 (11), 277.0 (18), 276.1 (15), 263.1 (10). IR (neat) $\nu_{max}$: 3039, 1315, 1292, 1138, 1092, 949, 891, 833, 783, 714, 683 cm$^{-1}$.

\subsection{1,24-dithia[11]helicene (endo-dithia[11]helicene) (7)}

This compound was prepared following the previous procedure, but using 3,6-bis((E)-2-(naphtho[1,2-b]tiophen-8-yl)vinyl)phenanthrene (\textbf{6}) as starting reagent. The residue was purified by column chromatography on silica gel (hexane-CH$_2$Cl$_2$ 8:2) to obtain a yellow solid. The product was recrystallized in CH$_2$Cl$_2$ to obtain yellow crystals.

\begin{figure}[H]
\centering
\includegraphics[width=.2\linewidth]{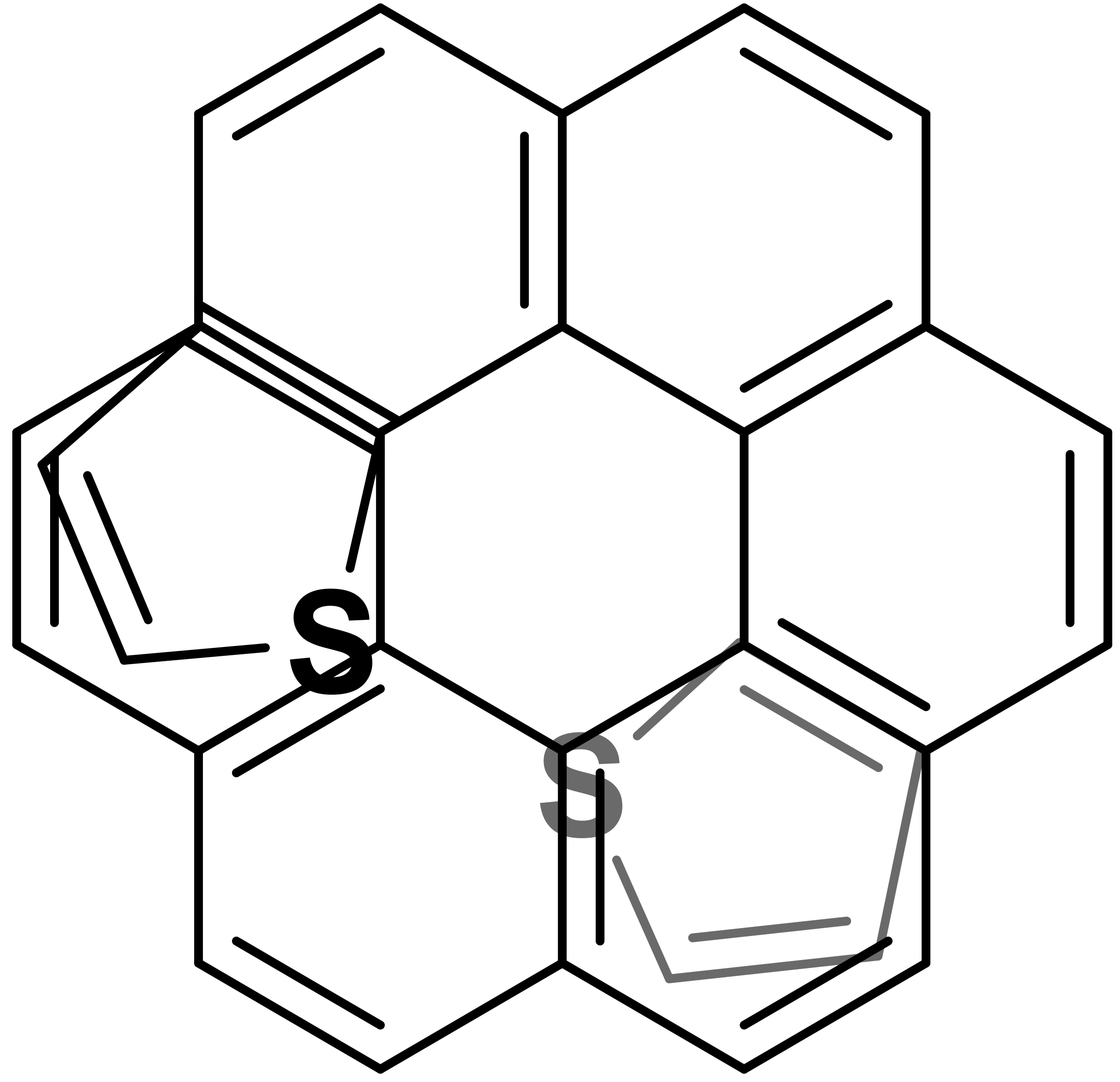}
\end{figure}

Yellow crystals in 46\% yield; R$_f$= 0.52 (hexane-CH$_2$Cl$_2$ 6:4); 1H-NMR (CDCl$_3$, 300MHz): $\delta$= 7.72 (s, 2H), 7.46 (d, J= 8.2 Hz, 2H), 7.44 (d, J= 8.4 Hz, 2H), 7.39 (d, J= 8.4 Hz, 2H), 7.30 (d, J=8.4 Hz, 2H), 7.24 (d, J= 8.2 Hz, 2H), 7.23 (d, J= 8.3 Hz, 2H), 7.21 (s, 4H), 6.83 (d, J= 5.4 Hz, 2H), 6.63 (d, J= 5.4 Hz, 2H) ppm. $^{13}$C-NMR (CDCl$_3$, 101 MHz): $\delta$=137.20 (-C-, 2C), 135.32 (-C-, 2C), 132.86 (-C-, 2C), 131.39 (-C-, 2C), 130.88 (-C-, 2C), 130.05 (-C-, 2C), 127.33 (-C-, 2C), 127.03 (-CH-, 2H), 127.00 (-CH-, 2C), 126.29 (-CH-, 2C), 126.19 (-CH-, 2C), 125.92 (-CH-, 2C), 125.69 (-CH-, 2C), 125.61 (-CH-, 2C), 124.69 (-CH-, 2C), 124.37 (-C-, 2C), 123.12 (-CH-, 2C), 122.73 (-C-, 2C), 122.72 (-C-, 2C), 121.56 (-CH-, 2C), 120.92 (-CH-, 2C) ppm. MS (EI, DIP) m/z: 592.2 (M$^+$+2, 19), 591.1 (M$^+$+1, 45), 590.2 (M$^+$, 100), 295.1 (23), 277.2 (11), 270.7 (12), 245.1 (7). IR (neat) $\nu_{max}$: 3039, 1300, 1265, 949, 833, 752, 717, 679 cm$^{-1}$.

\subsection{3,22-dithia[11]helicene} 

\textbf{$^1$H-NMR (CDCl$_3$, 400 MHz)}
\begin{figure}[H]
\centering
\includegraphics[width=.8\linewidth]{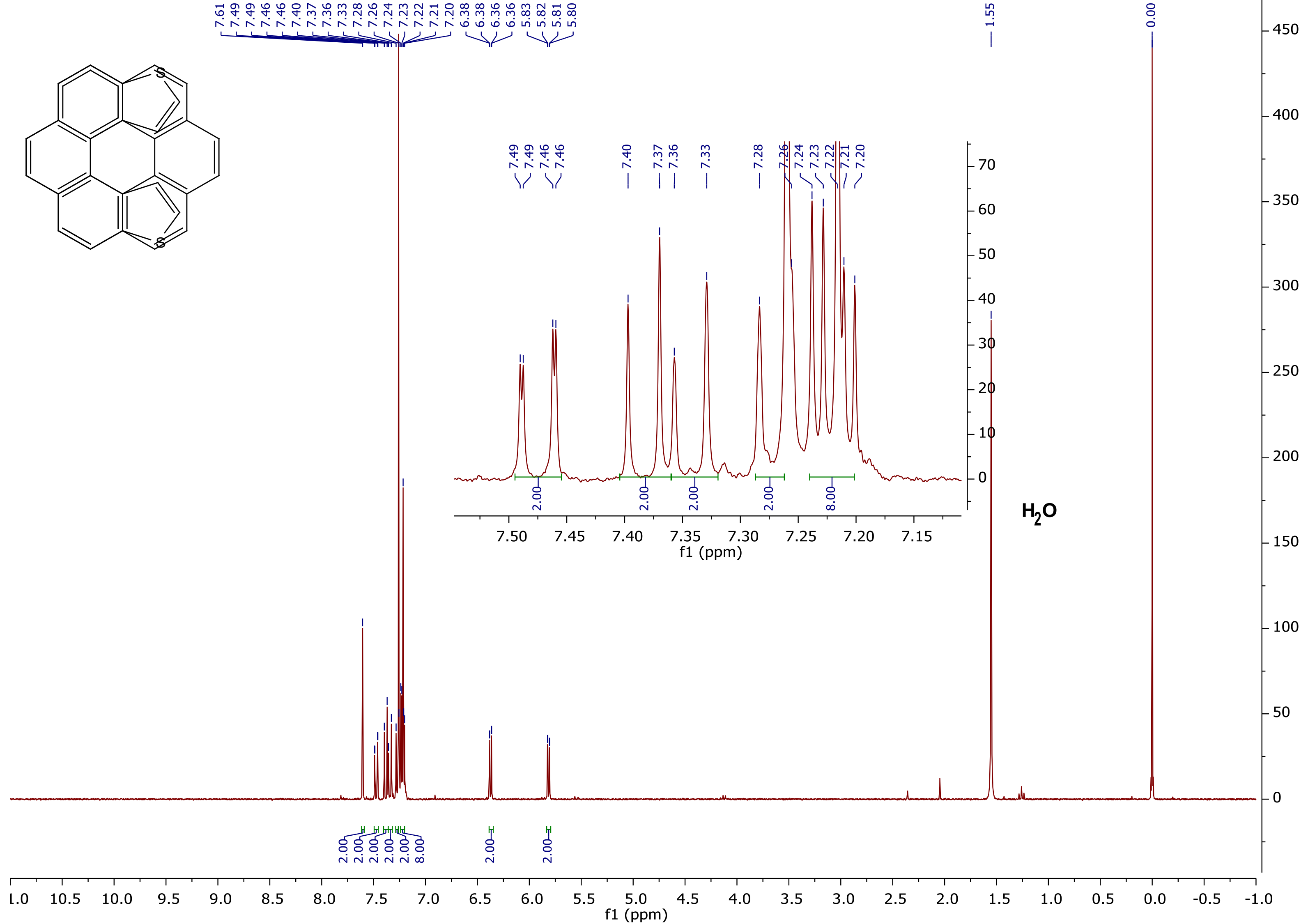}
\end{figure}

\textbf{$^{13}$C-NMR (CDCl$_3$, 101 MHz)}
\begin{figure}[H]
\centering
\includegraphics[width=.8\linewidth]{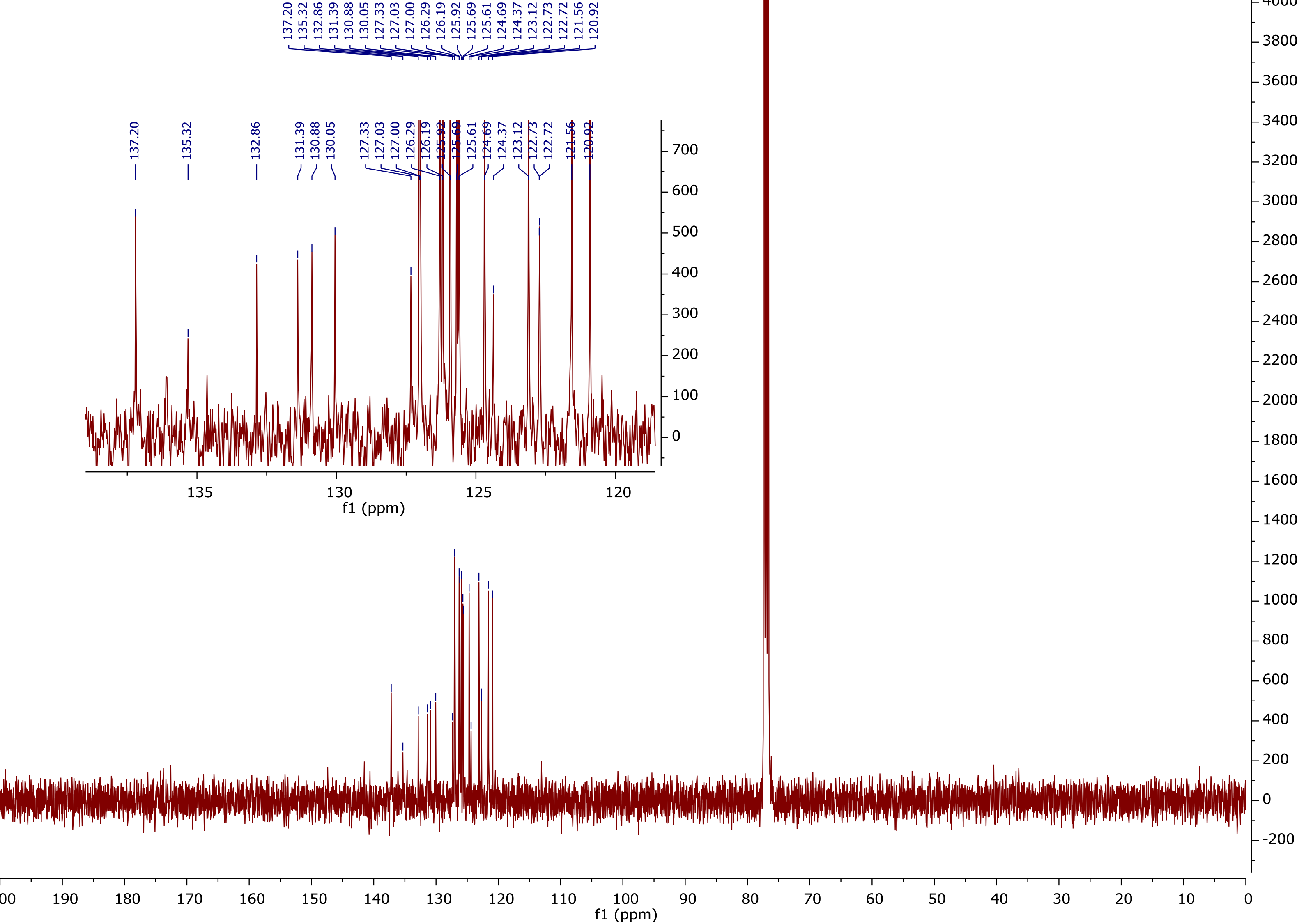}
\end{figure}

\subsection{1,24-dithia[11]helicene} 

\textbf{$^1$H-NMR (CDCl$_3$, 300 MHz)}
\begin{figure}[H]
\centering
\includegraphics[width=.8\linewidth]{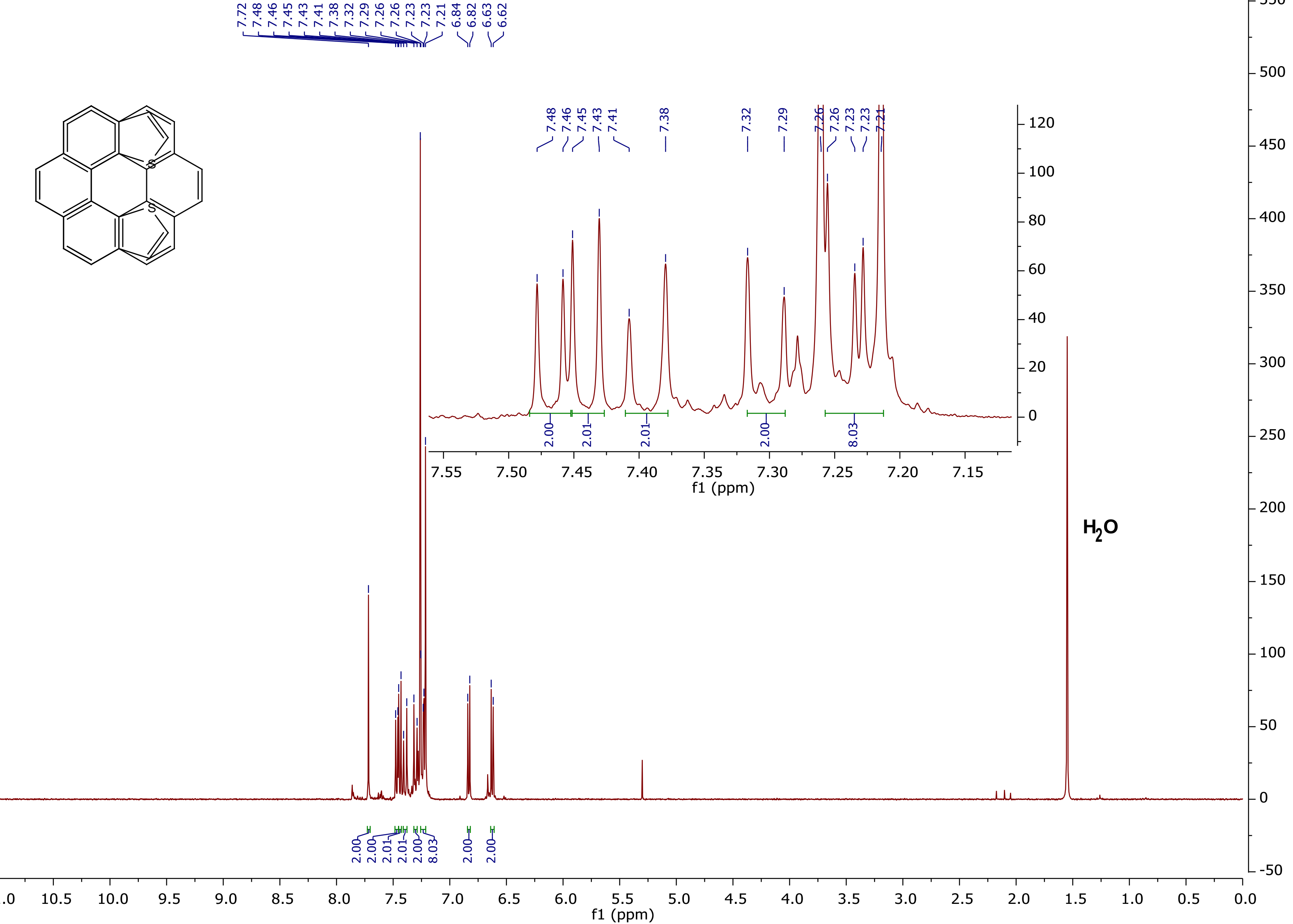}
\end{figure}

\textbf{$^{13}$C-NMR (CDCl$_3$, 101 MHz)}
\begin{figure}[H]
\centering
\includegraphics[width=.8\linewidth]{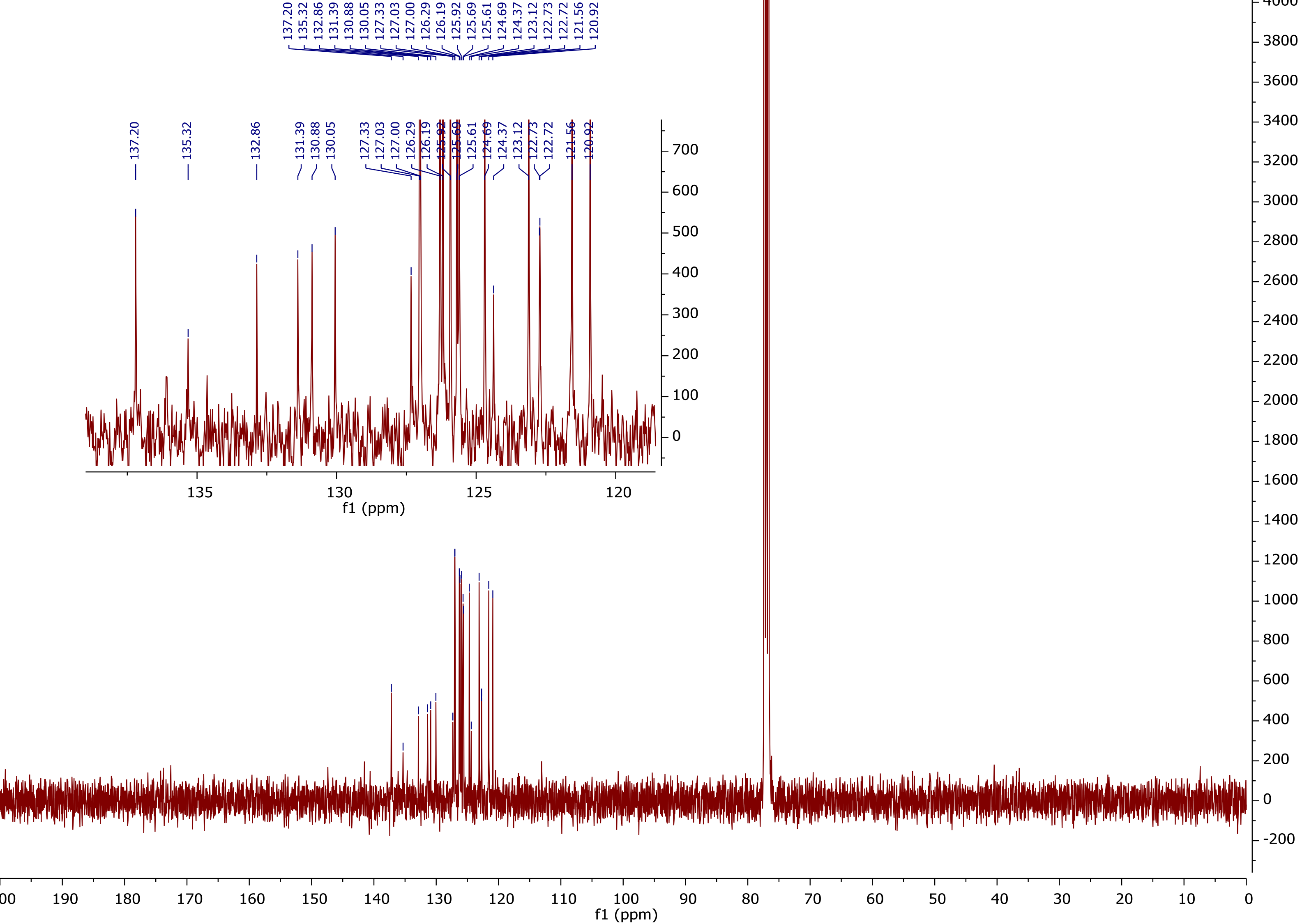}
\end{figure}

\subsection{2D histograms of the raw data of all helicenes measured at 0.1 V}

\begin{figure}[H]
\centering
\renewcommand\thefigure{S3}  
\includegraphics[width=1\linewidth]{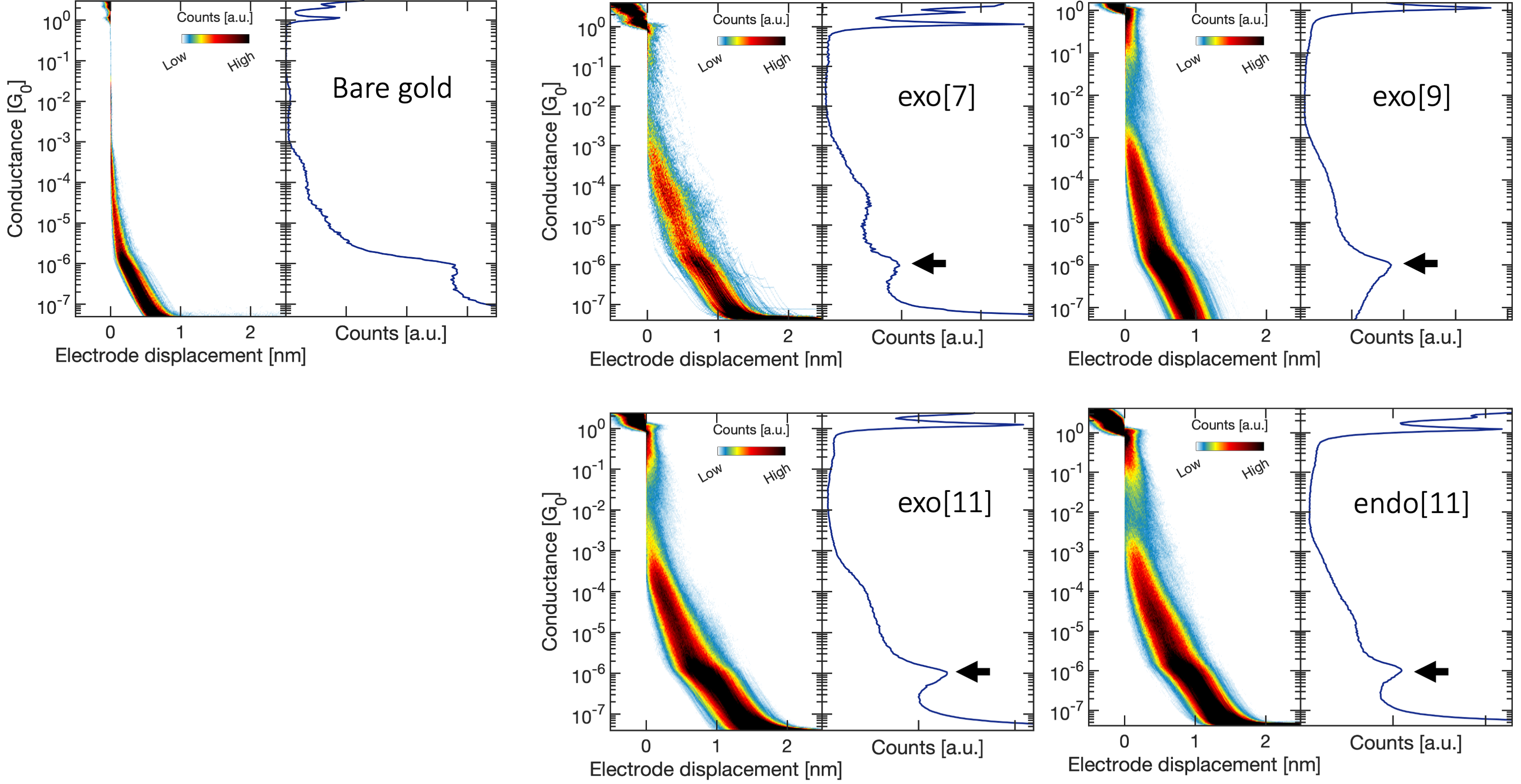}
\caption{2D histograms collecting all the raw data measured using MCBJ of bare gold and gold with helicenes.}
\label{fig:rawdata}
\end{figure}

\subsection{Evolution of the peak related to the log-amplifier with the bias voltage}

\begin{figure}[H]
\centering
\renewcommand\thefigure{S4} 
\includegraphics[width=1\linewidth]{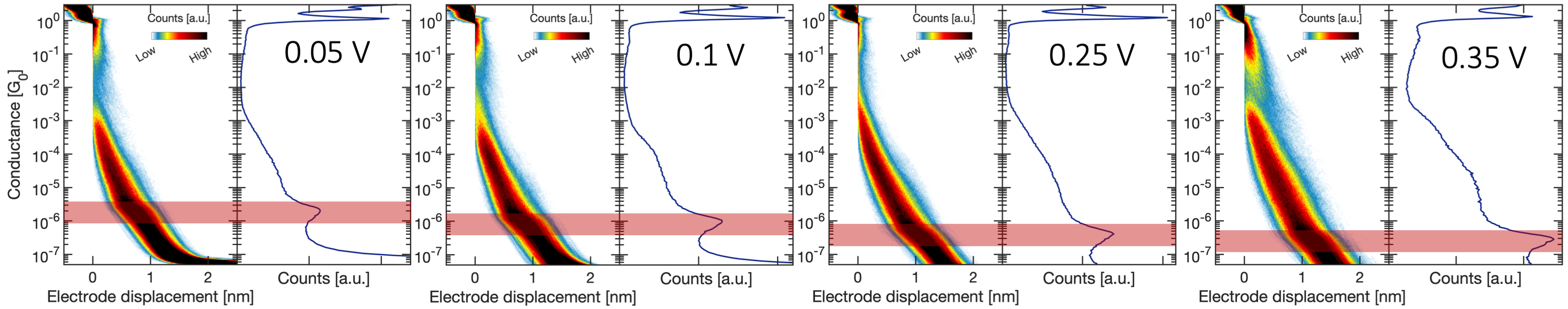}
\caption{Raw data obtained via MCBJ of exo[11]helicenes at different bias voltages. The peak related to the amplifier artifact goes down in value when increasing the bias voltage. The evolution is highlighted using a red-shaded area.}
\label{fig:artifact}
\end{figure}

\subsection{Details for running the clustering technique}

The input for the clustering considers a 2D histogram of each trace built by a matrix of 30x30 bins which is flatten by concatenating its rows. For this work, the algorithm focuses on the region between 10$^{-2}$ to 10$^{-5}$ G${_0}$, where molecular signatures typically appear, avoiding the region where the tunneling is the main contribution. The image is also built considering the range from 0 to 2 nm in displacement. Additionally, the 1D histogram is appended as input with 100 bins and limits -2 to -5 log(G/G$_0$). 

\begin{figure}[H]
\centering
\renewcommand\thefigure{S5} 
\includegraphics[width=1\linewidth]{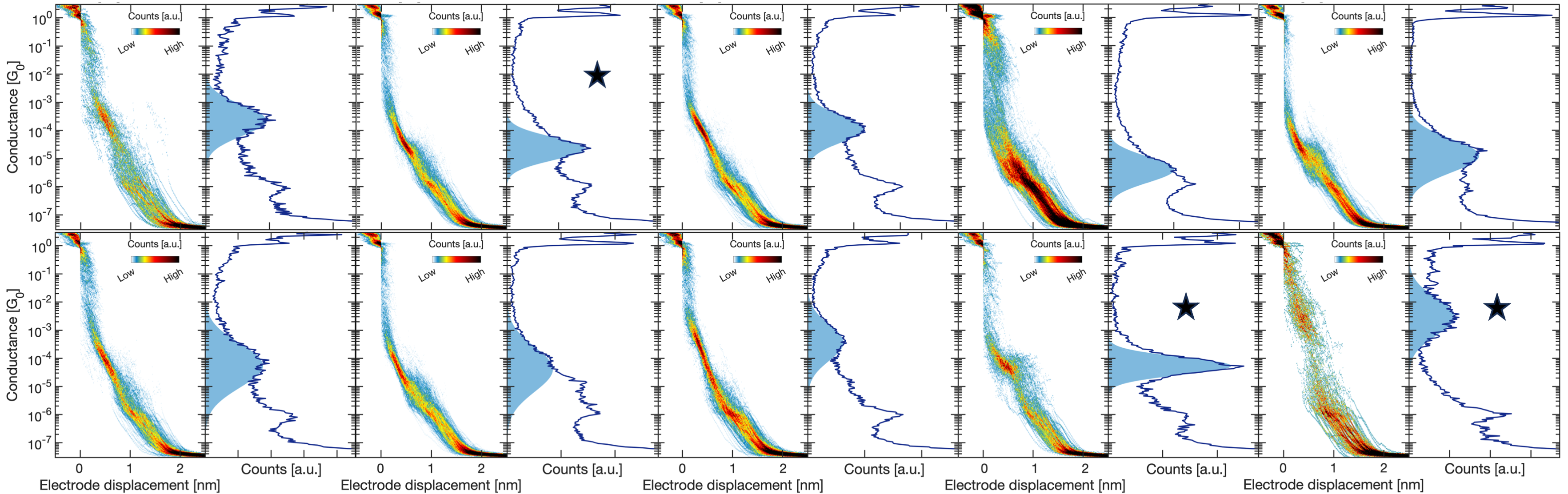}
\caption{2D/1D histograms of all the clusters obtained for the case of exo[11] helicenes measured at 0.1V. Marked clusters are the ones selected as examples in Figure 2 of the main text. }
\label{fig:clusterexo11}
\end{figure}

This input offers the details required for clustering the breaking traces into the predefined number of clusters. For the analysis of the data presented, the classification was performed using 3 clusters for the exo[7], and 10 clusters for the exo[9,11] and endo[11]. The output for the exo[11] is presented in Figure \ref{fig:clusterexo11}. Each 2D histogram is presented along the corresponding 1D histogram. Blue shaded area shows the Gaussian fit. Table \ref{tab:logConductance_100mV} collects the mean conductance values obtained.

\begin{table}[H]
\renewcommand\thetable{S\arabic{table}}  
\caption{Logarithmic fitted conductance values ($\mu$) and logarithmic deviations ($\sigma$) of the main assignments of the helicenes.}
    \centering
    \begin{tabular}{cccc}
    \hline
        exo[7]  & exo[9] & exo[11] & endo[11]\\
        \hline
                            & -2.5 $\pm$ 0.2     & -2.4 $\pm$ 0.2        & -2.5 $\pm$ 0.3    \\
                            & -3.31 $\pm$ 0.09   & -3.15 $\pm$ 0.08      & -3.14 $\pm$ 0.03  \\
        -4.19 $\pm$ 0.08    & -4.3 $\pm$ 0.3     & -4.4 $\pm$ 0.2        & -4.4 $\pm$ 0.2    \\
        -4.56 $\pm$ 0.07    & -4.8 $\pm$ 0.4     & -4.83 $\pm$ 0.01      & -4.82 $\pm$ 0.02  \\
    \hline
    \end{tabular}
    \label{tab:logConductance_100mV}
\end{table}

\begin{figure}[H]
\centering
\renewcommand\thefigure{S6} 
\includegraphics[width=0.5\linewidth]{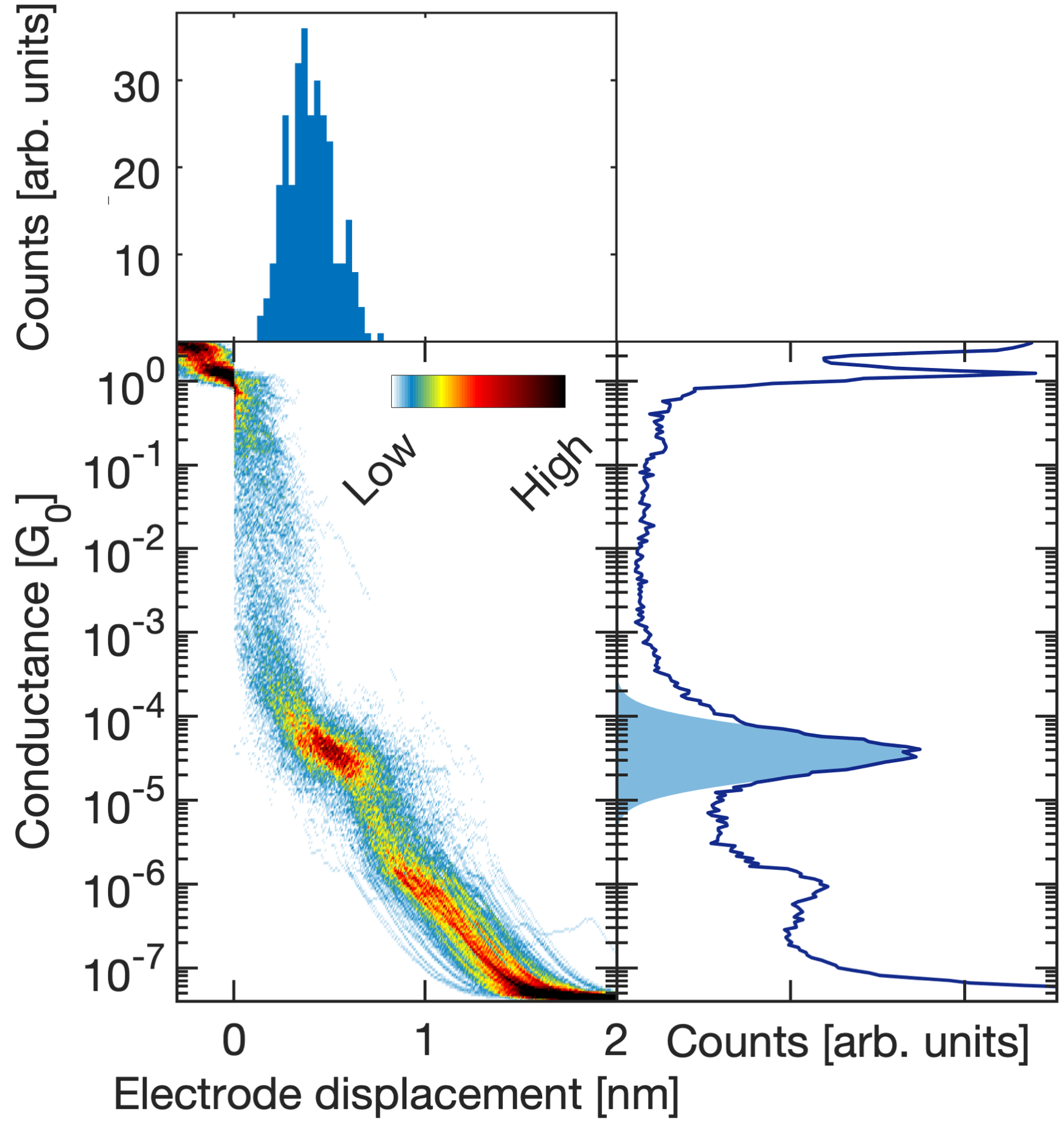}
\caption{2D/1D histogram of the conductance along 1D histogram of the length.}
\label{fig:length histogram}
\end{figure}

From the Gaussian fit, a length histogram can be built as to obtain an estimation of the most repeated plateau length of each cluster. An example of the composition of a 2D/1D histograms is shown in Figure \ref{fig:length histogram} along the length histogram built for that cluster.

\subsection{Cluster evolution with the bias voltage}
\begin{figure}[H]
\centering
\renewcommand\thefigure{S7} 
\includegraphics[width=1\linewidth]{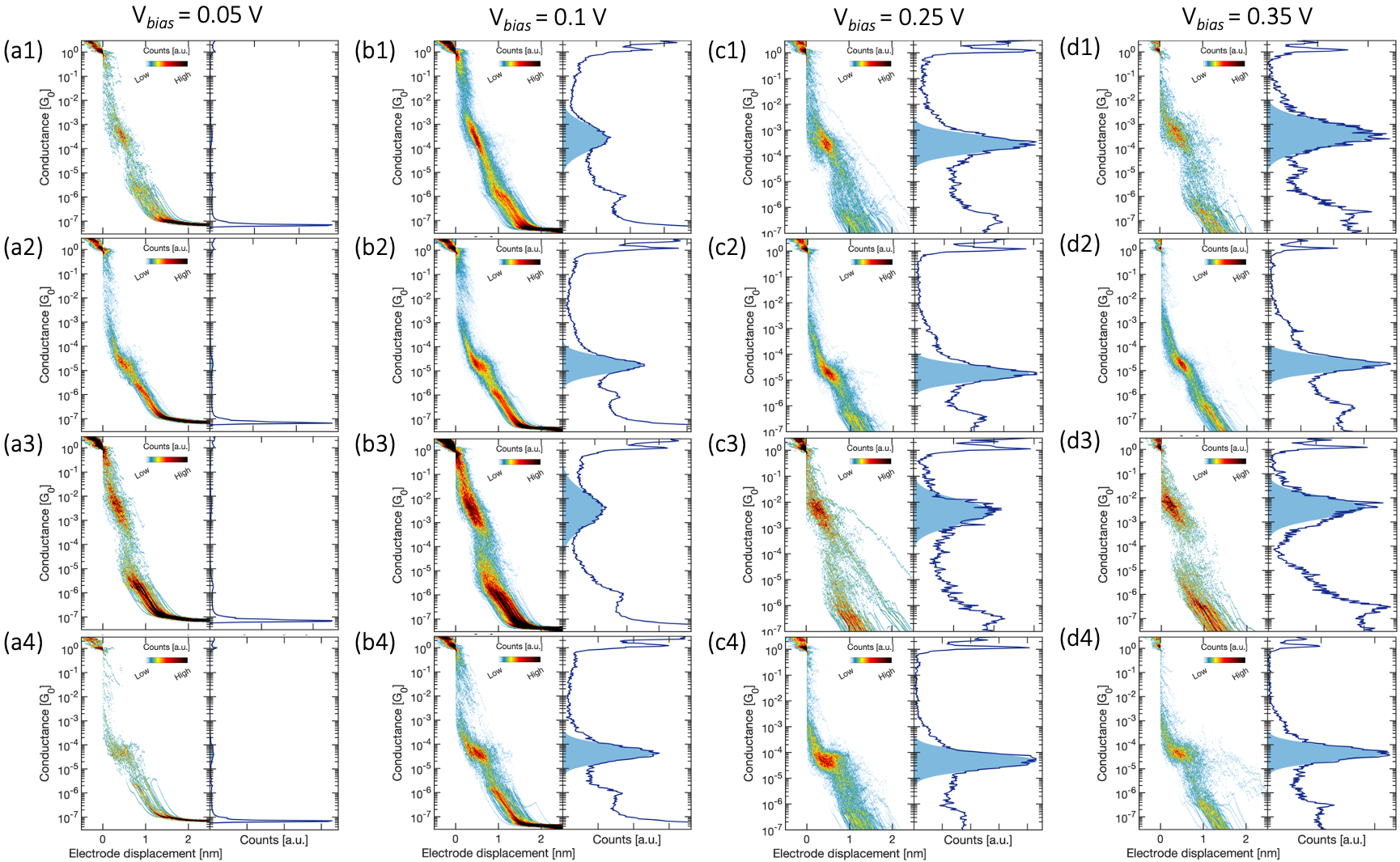}
\caption{Examples to observe the cluster evolution with the bias voltage for the endo[11] using 10 clusters.}
\label{fig:followcluster-endo11}
\end{figure}

\begin{figure}[H]
\centering
\renewcommand\thefigure{S8} 
\includegraphics[width=1\linewidth]{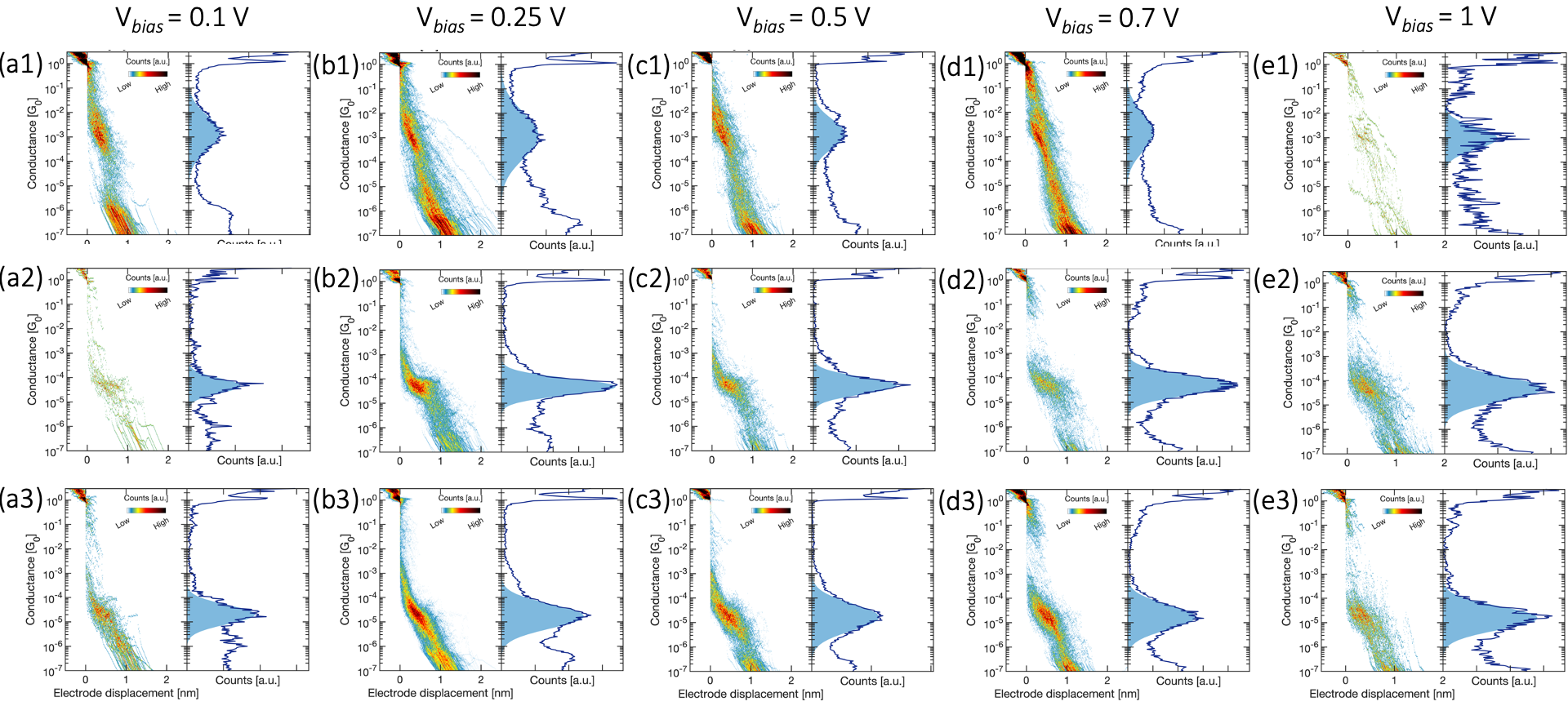}
\caption{Examples to observe the cluster evolution with the high bias voltage for the endo[11] using 10 clusters.}
\label{fig:followcluster-endo11high}
\end{figure}

As reference, Figures \ref{fig:followcluster-endo11} and \ref{fig:followcluster-endo11high} display the common clusters across different bias voltages.

\subsection{Density functional theory details and transmission curves}
Before conducting the conductance calculations, a geometry optimization is performed for the electrode-molecule system. For our calculations, HSE06 functional was employed as it is widely used in metal-organic systems \cite{HSEH1PBE03, HSEH1PBE04, camarasagómez2023transferable, JanusMATyHSE}. The conductance calculations were performed using ANT.GAUSSIAN code \cite{palacios2001fullerene, palacios2002transport, ANTG}, which is built upon GAUSSIAN09. For specific details, refer to the main text and supplementary materials of reference \cite{DednamACS2023}.

\begin{figure}[H]
\centering
\renewcommand\thefigure{S9} 
\includegraphics[width=.9\linewidth]{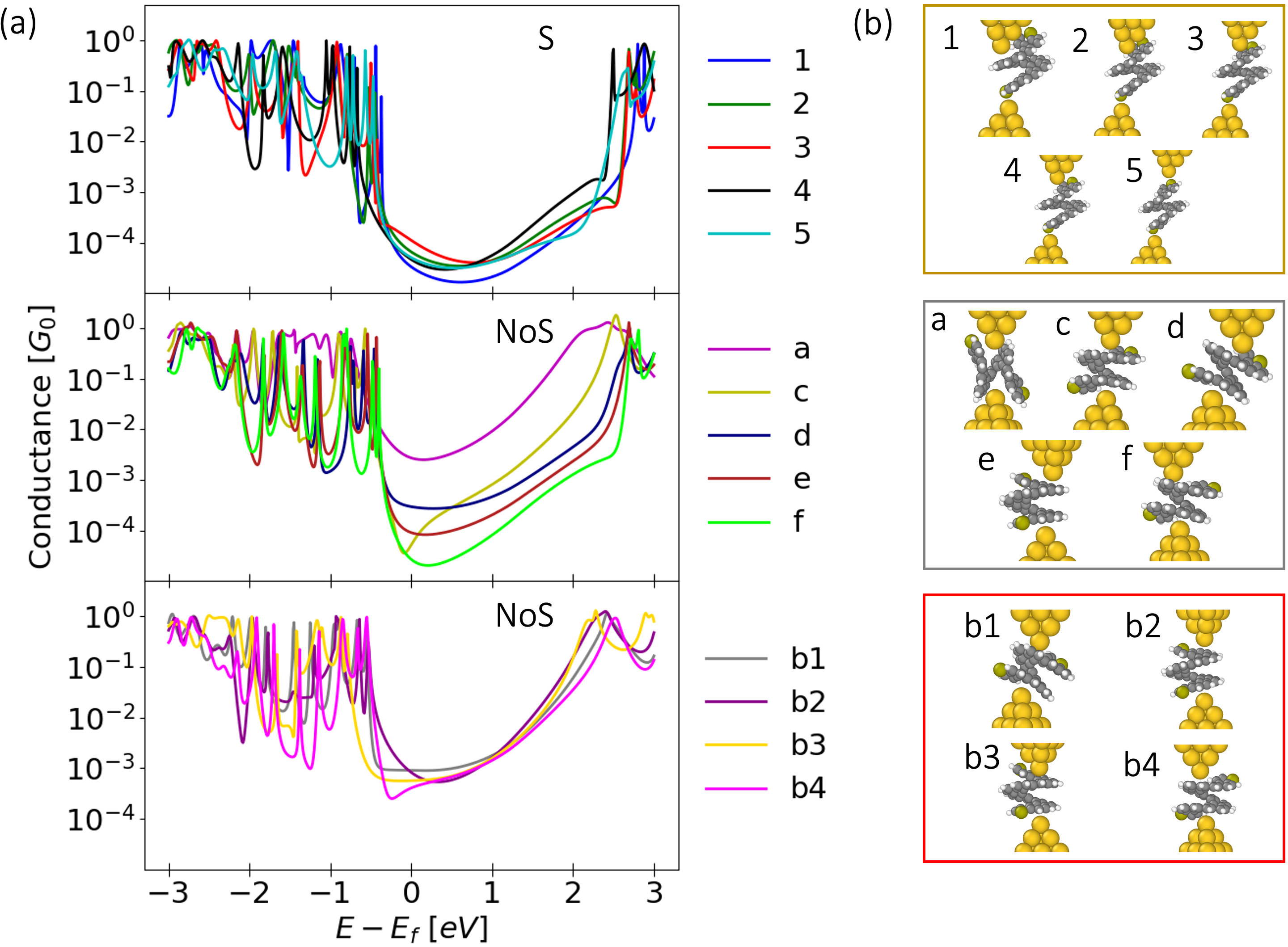}
\caption{Panel (a) displays conductance curves obtained as a function of energy. The upper section features molecules linked by sulfur, the middle and bottom showcases scenarios without sulfur connections. The curves exhibit similar patterns within each group. Panel (b) provides illustrations of the legends from the previous panel.}
\label{fig:DFTillustrations}
\end{figure}

The computed curves of transmission, in terms of conductance ($G_{0}$) as a function of energy, are shown in a vertically stacked graph in Figure \ref{fig:DFTillustrations}(a). The upper panel corresponds to molecules that are linked by at least one sulfur atom (labelled 1-5). In contrast, the middle and bottom panels show scenarios where the molecules are connected without any sulfur (labelled a-f and b1-b-4). The latter ones are marked by a red circle in Figure 4(b) of the main text, as they correspond to a cluster of points of configurations with similar conductance values. Additionally, panel (b) illustrates the legends corresponding to the aforementioned panel. It can be observed that the three grouped curves exhibit similar shapes. In all cases, although the HOMO is close to the Fermi energy, it is out of resonance.

\bibliography{main.bib}